\documentclass[12pt]{article}
\usepackage[utf8]{inputenc}
\usepackage{fullpage,color}
\usepackage{amsmath}
\usepackage{amsthm}
\usepackage{amssymb}
\usepackage{latexsym}
\usepackage{natbib,booktabs}
\usepackage{setspace}
\usepackage{boxedminipage}
\usepackage{listings,url}
\usepackage[lined,boxed]{algorithm2e}
\usepackage[lined]{algorithm2e}

\usepackage{ifpdf}

\ifpdf
\usepackage[pdftex]{graphicx}
\else
\usepackage{graphicx}
\fi

\newcommand\bs{\boldsymbol}
\newcommand{\bmu}{\mbox{\boldmath{$\mu$}}}
\newcommand{\bth}{\mbox{\boldmath{$\theta$}}}
\newcommand{\balp}{\mbox{\boldmath{$\alpha$}}}
\newcommand{\blam}{\mbox{\boldmath{$\lambda$}}}

\newcommand{\bphi}{\mbox{\boldmath{$\phi$}}}
\newcommand{\bhatphi}{\mbox{\boldmath{$\hat\phi$}}}
\newcommand{\bnu}{\mbox{\boldmath{$\nu$}}}
\newcommand{\deuc}{d_2}
\newcommand{\dmin}{d_p}
\newcommand{\dcos}{d_{\rm cos}}
\newcommand{\Dcos}{D_{\rm cos}}
\newcommand{\dhell}{d_{\rm H}}
\newcommand{\nclus}{k}
\newcommand{\ntemp}{n_T}

\newcommand{\bx}{\mbox{\boldmath{$x$}}}
\newcommand{\by}{\mbox{\boldmath{$y$}}}
\newcommand{\bz}{\mbox{\boldmath{$z$}}}

\newcommand{\br}{\mbox{\boldmath{$r$}}}

\newcommand{\bT}{\mbox{\boldmath{$T$}}}

\newcommand{\E}{\text{E}}

\newcommand{\rt}{\right}
\newcommand{\lt}{ \left}
\newcommand{\iid}{\buildrel{\rm iid}\over\sim}
\DeclareMathOperator*{\argmin}{arg\,min}

\newcommand{\iter}{^{(\ell)}}
\newcommand{\clone}{_{(1)}}
\newcommand{\cltwo}{_{(2)}}
\newcommand{\clnc}{_{(\nclus)}}
\newcommand{\clj}{_{(c)}}

\newcommand{\edit}[1]{{{#1}}}

\title{\bf
\edit{Preprocessing Solar Images while Preserving their Latent Structure}
}
\author{Nathan M. Stein \\ \normalsize Department of Statistics \\ \normalsize University of Pennsylvania
\and 
David A. van Dyk \\ \normalsize Statistics Section\\ \normalsize Imperial College London
\and 
Vinay L. Kashyap \\ \normalsize High-Energy Astrophysics Division \\ \normalsize Harvard Smithsonian Center for Astrophysics}

\begin{document}

\ifpdf
\DeclareGraphicsExtensions{.pdf, .jpg, .tif}
\else
\DeclareGraphicsExtensions{.eps, .jpg}
\fi

\maketitle
\abstract{Telescopes such as the Atmospheric Imaging Assembly aboard the Solar Dynamics Observatory, a NASA satellite, collect massive streams of high resolution images of the Sun through multiple wavelength filters. Reconstructing pixel-by-pixel thermal properties based on these images can be framed as an ill-posed inverse problem with Poisson noise, but this reconstruction is computationally expensive and there is disagreement among researchers about what regularization or prior assumptions are most appropriate. 
This article presents an image segmentation framework for preprocessing such images in order to reduce the data volume while preserving as much thermal information as possible for later downstream analyses. The resulting segmented images reflect thermal properties but do not depend on solving the ill-posed inverse problem.
\edit{This allows users to avoid the Poisson inverse problem altogether or to tackle it on each of $\sim$10 segments rather than on each of $\sim$10$^7$ pixels, reducing computing time by a factor of $\sim$10$^6$.}
We employ a parametric class of dissimilarities that can be expressed as cosine dissimilarity functions or Hellinger distances between nonlinearly transformed vectors of multi-passband observations in each pixel. We develop a decision theoretic framework for choosing the dissimilarity that minimizes the expected loss that arises when estimating identifiable thermal properties based on segmented images rather than on a pixel-by-pixel basis. We also examine the efficacy of different dissimilarities for recovering clusters in the underlying thermal properties. The expected losses are computed under scientifically motivated prior distributions. Two simulation studies guide our choices of dissimilarity function. We illustrate our method by segmenting images of a coronal hole observed on 26 February 2015.
}




\section{Solar image segmentation}
\label{sec:solar}
\subsection{Tracing the differential emission measure} 

The solar {\it corona} is the region of the Sun's atmosphere furthest from its surface; it consists of hot plasma that is more than about $10^6$ km above the surface. One of the major unsolved questions in coronal astrophysics is the mechanism by which the energy stored in magnetic fields is transferred to the plasma to heat the corona to its observed ultra high temperatures ($10^6$--$10^7$ Kelvin), millions of degrees hotter than at the surface of the Sun. Understanding such mechanisms is important not just in astronomy: solar activity directly affects Earth's climate, and solar flares and coronal mass ejections can seriously affect both ground- and spaced-based electronic infrastructure \citep[e.g.,][]{oden:01,bold:02}.

\begin{figure}[htbp]
\includegraphics[width=0.34\textwidth]{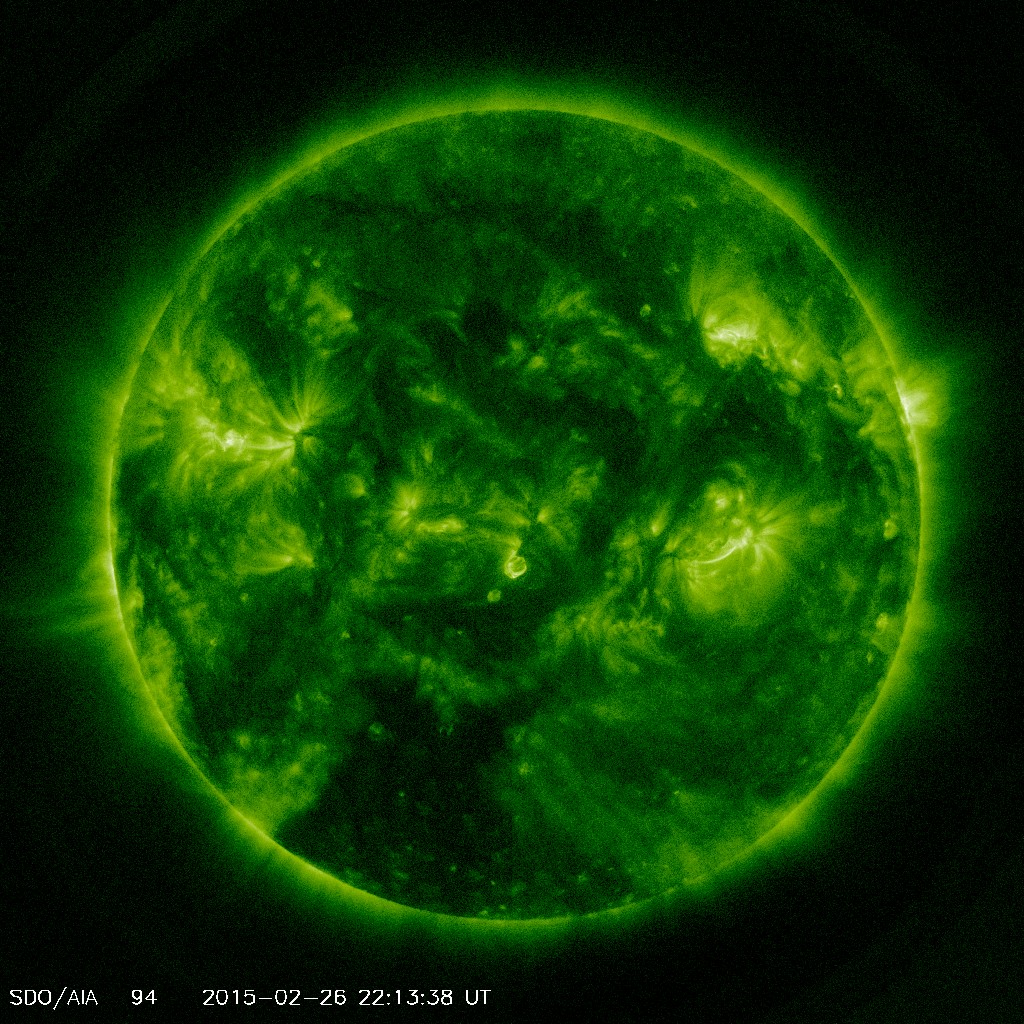}
\kern -2mm\includegraphics[width=0.34\textwidth]{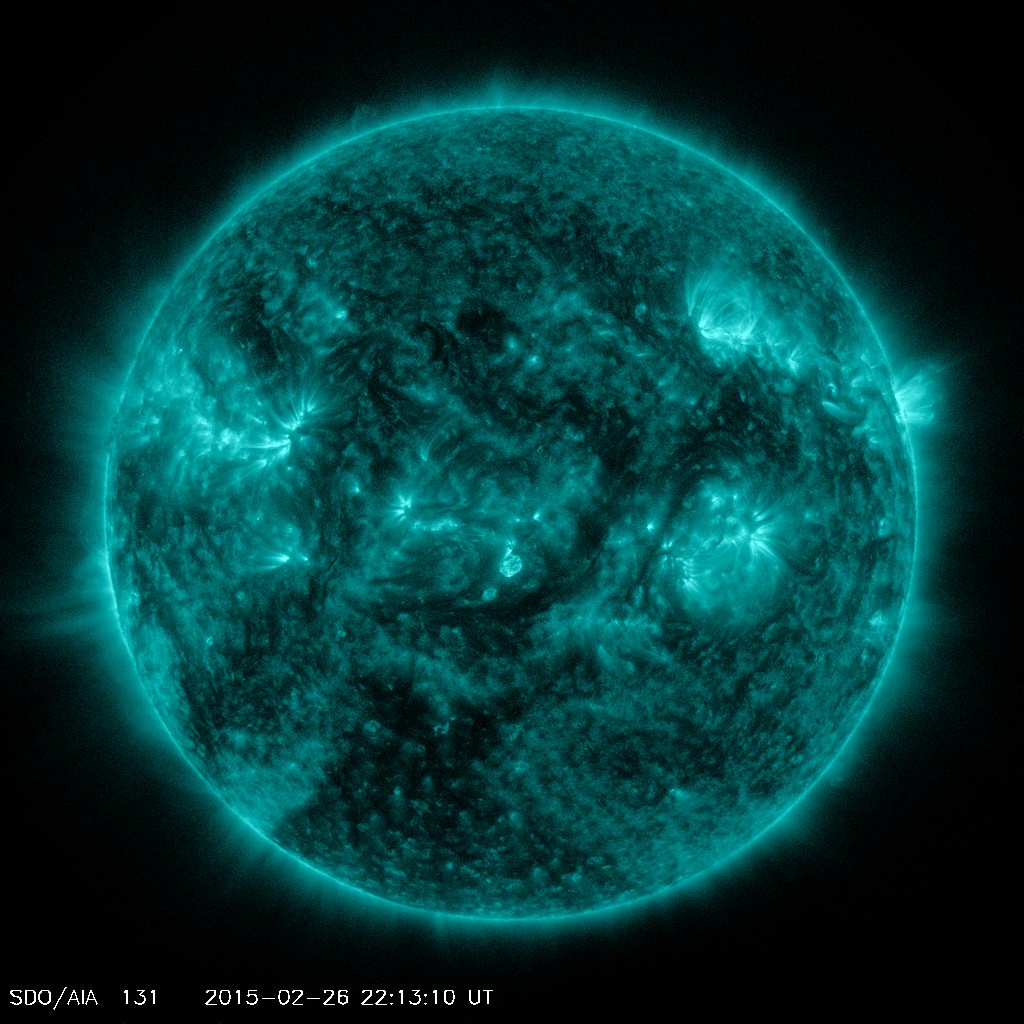}
\kern -2mm\includegraphics[width=0.34\textwidth]{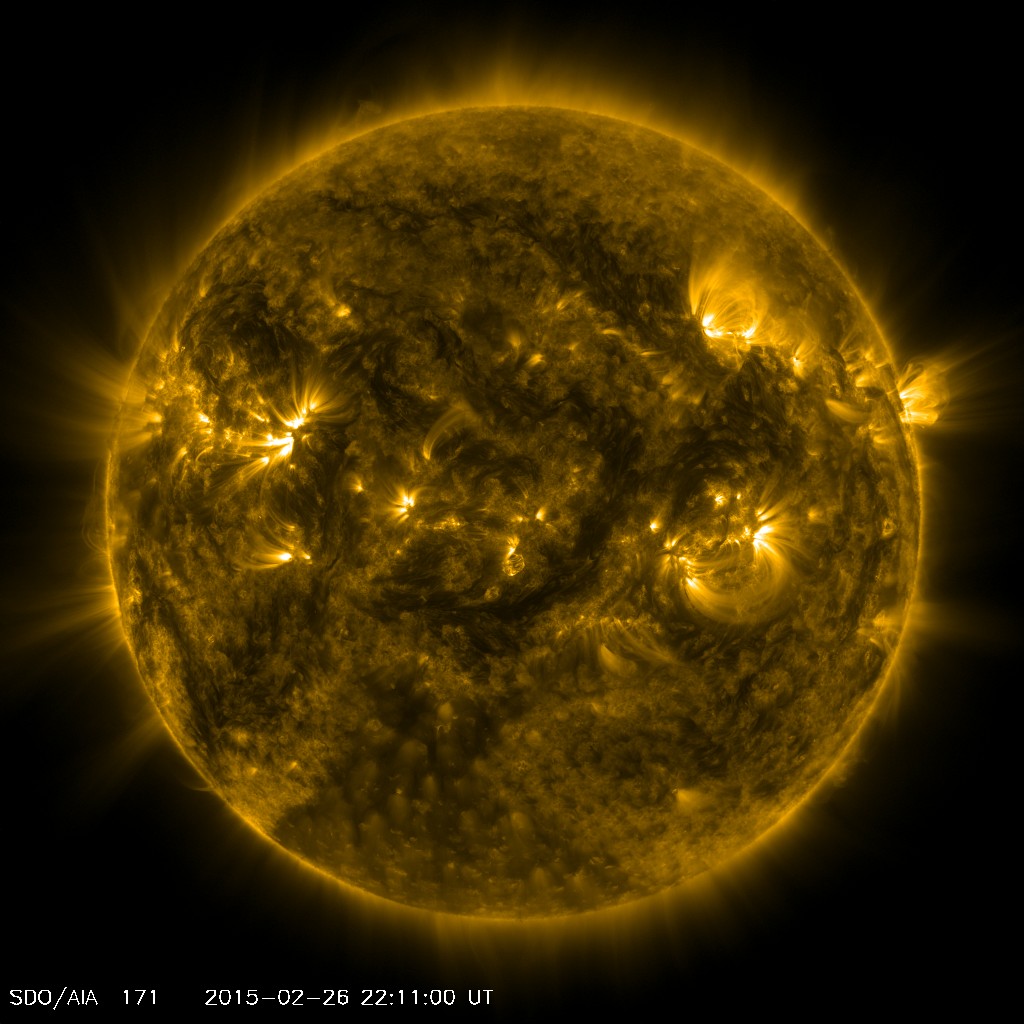}

\kern -1mm
\includegraphics[width=0.34\textwidth]{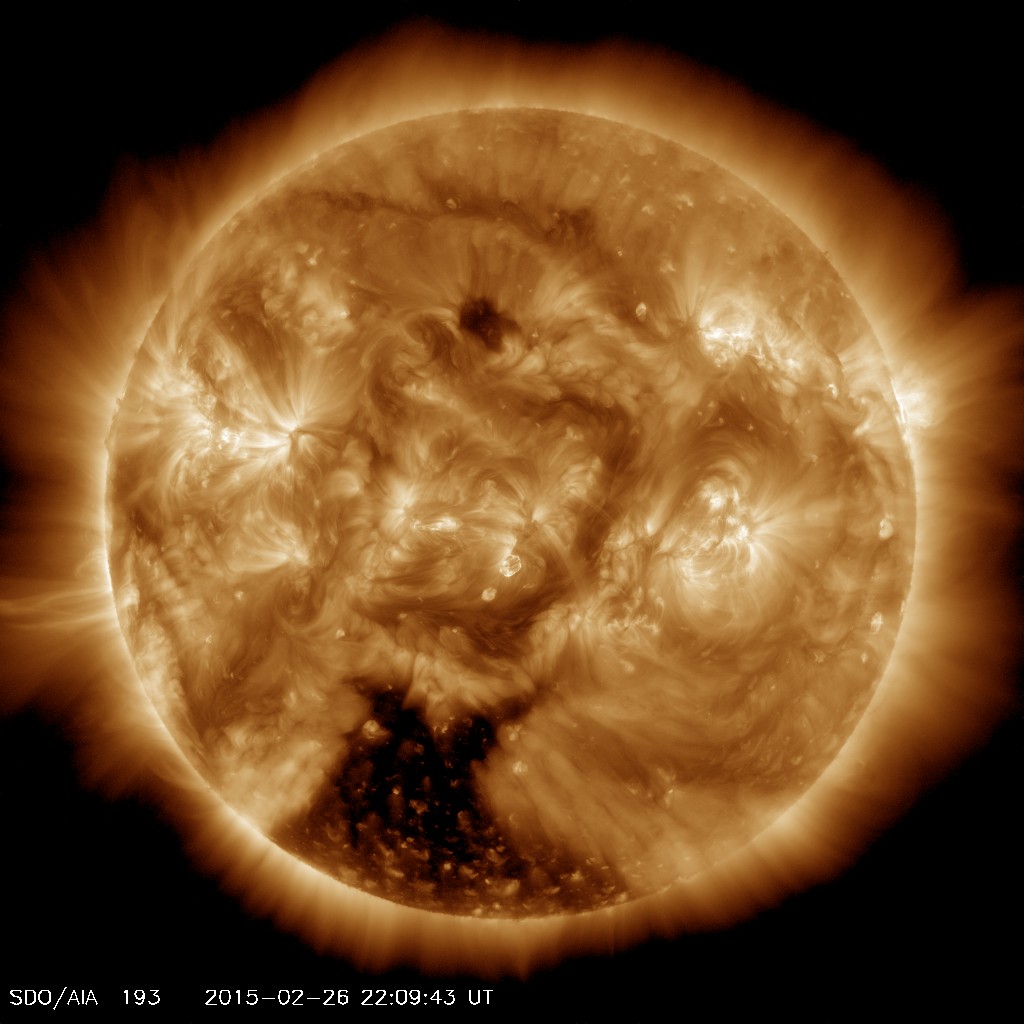}
\kern -2mm\includegraphics[width=0.34\textwidth]{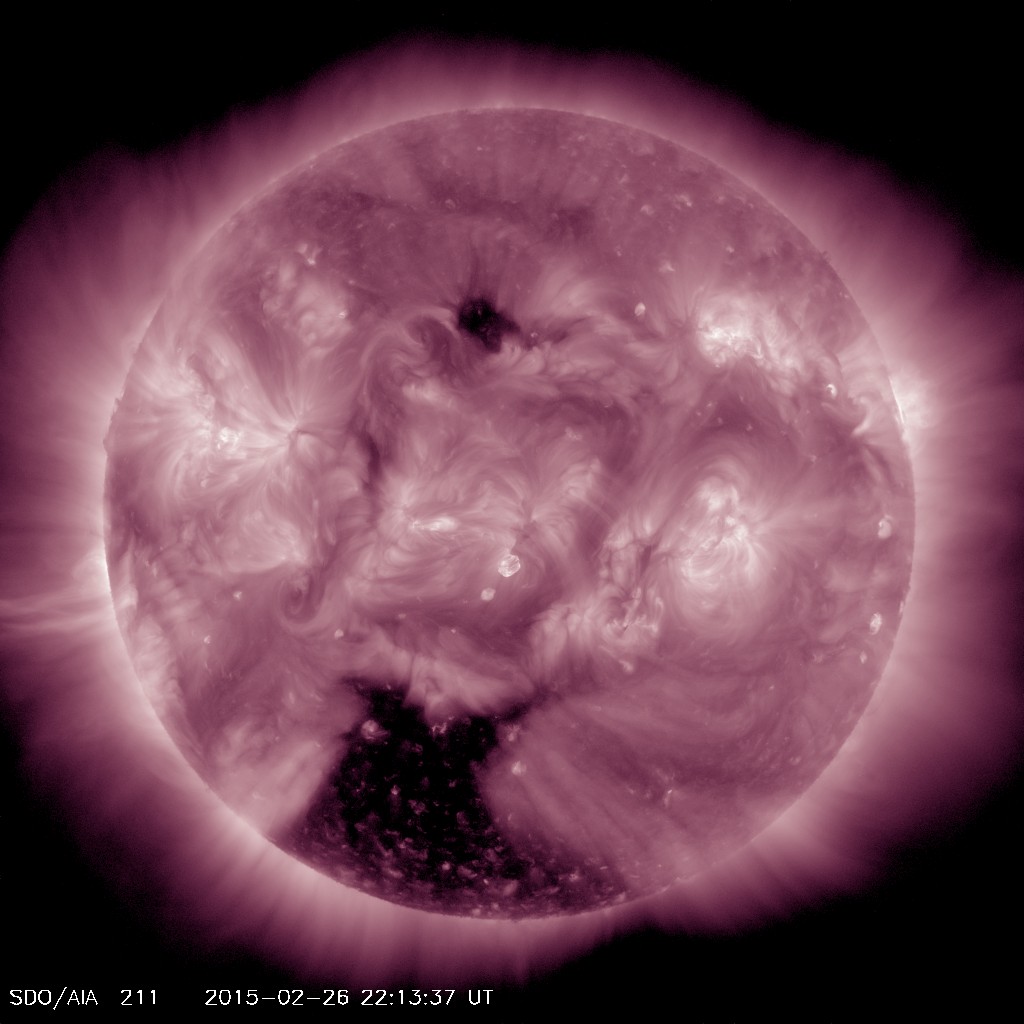}
\kern -2mm\includegraphics[width=0.34\textwidth]{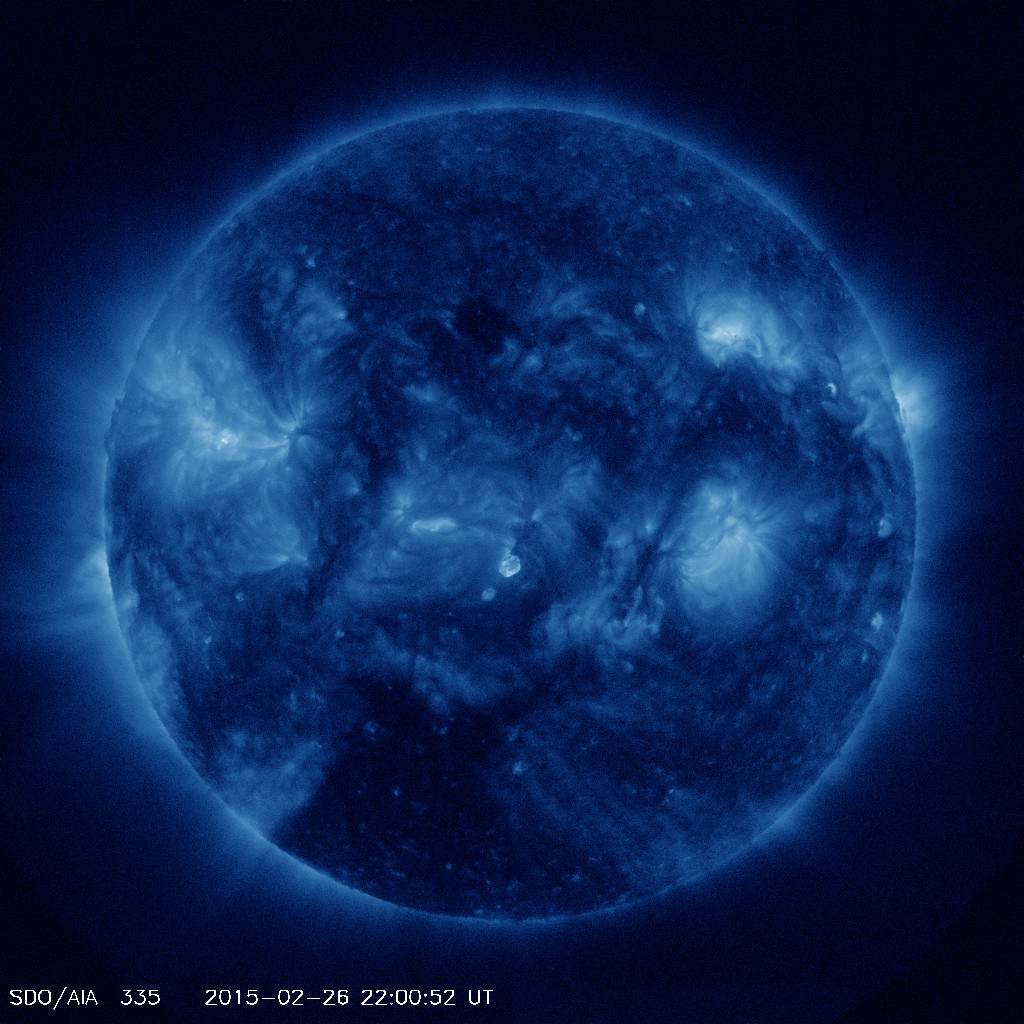}\\
\caption{An image of the Sun taken on 26 February 2015 in six of the seven extreme ultraviolet AIA passbands. Each passband records a 4096$\times$4096 pixel image of photon counts; higher image intensity represents higher photon counts. The colors are artificial, but emphasize the differing wavelength ranges in each filter.
The first row correspond to the passbands centered at 94\AA, 131\AA, and 171\AA~and
the second row to 193\AA, 211\AA, 
and 335\AA; ten Angstroms (\AA) is one nanometer. (Courtesy of NASA/SDO and the AIA, EVE, and HMI science teams.)
}
\label{fig:data}
\end{figure}

To study the solar corona, astronomers monitor the Sun closely and nearly continuously using space-borne platforms like {\sl Hinode} and the {\sl Solar Dynamics Observatory}\footnote{\url{http://sdo.gsfc.nasa.gov}} (SDO).  The Atmospheric Imaging Assembly (AIA) aboard the SDO is a four-telescope array that obtains snapshots of the solar corona in seven different extreme ultraviolet and soft X-ray filters; the distribution of electromagnetic wavelengths recorded by each filter is called its  {\it passband}.   The snapshots are obtained at the very high cadence rate of approximately one single-passband 4096$\times$4096 image every second; Figure~\ref{fig:data} illustrates a single snapshot in six of the seven passbands. Future observatories will offer even more impressive imaging. The (optical) Daniel K. Inouye Solar Telescope is currently under construction in Hawaii and is expected to resolve solar features 30km in diameter. 


To manage these massive data streams, new automated feature recognition and tracking methods are critically needed, especially those that account for underlying physical processes, such as thermal properties. The ability to reliably isolate and track thermal structures in the solar corona may provide strong constraints on the theoretical descriptions of the emergence of magnetic structures, loop formation, coronal heating, and other phenomena.

Temperature varies naturally across the solar corona, and its distribution is characterized by the so-called {\it differential emission measure} (DEM).  \edit{The DEM quantifies the amount of plasma present at a given temperature.%
\footnote{\edit{The DEM can be viewed as a density function of the plasma temperature; it quantifies the amount of plasma existing with temperature in each infinitesimal temperature bin. Formally, the DEM is the derivative of the {\it emission measure} with respect to plasma temperature (or log$_{10}$ temperature). 
The emission measure quantifies the amount of plasma that is emitting thermal X-ray or extreme ultraviolet photons, and is the product of the volume,  
the number density (particle count per unit volume) of electrons, and the 
number density of protons in that volume.}} 
We can define the DEM of the entire solar corona or of some subvolume. Each image pixel corresponds to a subvolume of the corona, specifically the massive three dimensional column of the corona extending from the surface of the Sun toward the telescope. A pixel's DEM characterizes the distribution of the temperature of the plasma in the coronal column corresponding to that pixel. 
This distribution, however, is unnormalized: a pixel's (temperature) integrated DEM is a measure of the total amount of plasma in that pixel.  

We are interested in how the DEM varies from pixel to pixel and in clusters of pixels with similar DEMs.  More precisely, to better understand the thermal structure and energetics of the corona,} 
we aim to identify coherent patterns of thermal activity in streams of multi-passband solar images, such as those produced by the AIA. Typically this corresponds to looking for regions with similar temperature distributions, that is regions with similar {\it normalized} DEMs. \edit{Thus, we aim to identify structure in how the normalized DEM varies across the corona.} Unfortunately, the observed passband data are only an imperfect proxy for the thermal characteristics of interest. Using them to accurately reconstruct thermal characteristics or to directly estimate the pixel DEMs poses significant challenges \citep[e.g.,][]{judg:etal:97}. \edit{As described in Section~\ref{sec:imgseg}, we propose a strategy that avoids such direct estimation.}

\begin{figure}[t]
\includegraphics[width=1.00\textwidth]{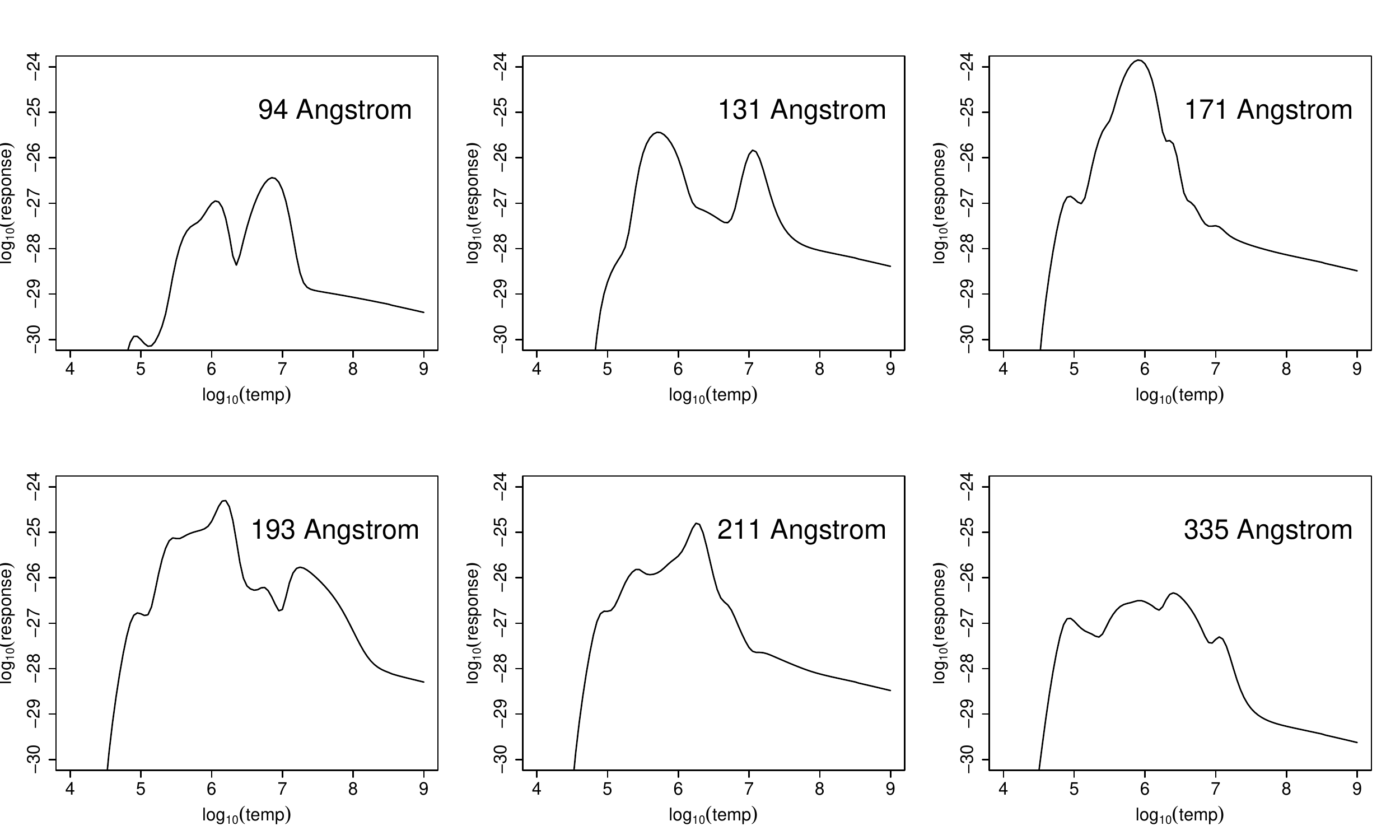}
\caption{The temperature response functions of the 6 AIA passbands for the images in Figure~\ref{fig:data}. Using a discretization of temperature, these response functions make up the rows of the matrix $\mathbf R$ in (\ref{eq:lin-inv}).}
\label{fig:Rplot}
\end{figure}

In this article, we formulate our notation and methodology in terms of images obtained with the AIA. In a multi-passband image, we observe each  pixel in $b$ passbands. We denote the resulting $b$ values in pixel $i$ by the vector  $\by_i = (y_{i1},\ldots,y_{ib})$ and the collection of pixels in a multi-passband image by the set of vectors $\{\by_1,\ldots,\by_n\}$, where  $n$ is the number of pixels.
 Each $y_{ij}$ is a count of the number of photons recorded by the detector. Each of the $b$ filters is characterized by a {\it temperature response function} 
describing the sensitivity of that filter to photons emitted from plasma at a given temperature.   
 In practice, temperature is discretized into $\ntemp$ bins so that the temperature response functions can be collected into a $b \times \ntemp$ matrix $\mathbf{R}$, where each row of $\mathbf{R}$ is the temperature response function for a particular passband; these are plotted in Figure~\ref{fig:Rplot}. Typically, the temperature is represented as a uniform grid in $\log_{10}(\hbox{degrees Kelvin})$ ranging from $5.5$ to $8$, with bin widths of $0.05$ or $0.1$.  Integrating the pixel-specific DEM over each of these temperature bins, we can summarize it with the $\ntemp\times 1$ vector $\bmu_i = (\mu_{i1}, \ldots, \mu_{i\ntemp})$.  The exposure times of the $b$ filters may differ and are tabulated in a diagonal matrix $\mathbf{A} = \text{diag}(\tau_1, \ldots, \tau_b)$, where $\tau_j$ is the exposure time in passband $j$. Given $\mathbf{R}$, $\mathbf{A}$, and $\bmu$, we can model the observed counts in pixel $i$, $\by_i$, as independent Poisson random variables with mean vector
\begin{equation}
\blam_i = \mathbf{A R } \bmu_i,
\label{eq:lin-inv}
\end{equation}
where $\blam_i = (\lambda_{i1}, \ldots, \lambda_{ib})$ is the vector of expected (multi-passband) counts in pixel $i$.
Typically, the thermal characteristics of interest (for example, the mean temperature), are functions only of $\bth_i = \bmu_i / \nu_i$, where $\nu_i = \sum_{j=1}^{\ntemp} \mu_{ij}$ is a nuisance parameter. Our scientific goal is to identify interesting spatial or spatiotemporal structure in the values of $\bth_i$. 
\edit{The standard strategy is to first fit (\ref{eq:lin-inv}) separately in each pixel and then to identify patterns in the fitted $\hat\bmu_i$. Because it is not feasible in practice to do this on $\sim$10$^7$ pixels (in each time frame), researchers focus on smaller subregions of the Sun.}

Poisson models with mean of the form given in (\ref{eq:lin-inv}) are quite common in practice (e.g., in image reconstruction problems such as positron emission tomography) and fitting them via maximum likelihood poses notorious computational difficulties, both because $\mathbf{R}$ may be ill-conditioned or singular and because $\bmu$ is constrained to be positive \citep[e.g.,][]{shep:vard:82,gree:90, esch:etal:04, bard:gold:09}.  In the best of cases, the likelihood tends to be highly multimodal and favor irregular choices of $\bmu$.
In our setting, such difficulties are exacerbated because the number of passbands is fewer than the desired number of bins in the reconstructed DEM. Thus, some sort of regularization is essential.

Although the choice of regularization can be quite influential in practice, there is no agreement as to what characteristics in the DEM should be encouraged by regularization (e.g., smoothness). The result is an array of methods for DEM reconstruction that can give different results. These methods range from forward-fitting splines ({\tt xrt\_dem\_iterative2.pro}, part of the Solar SoftWare package used to analyze solar data), to Chebyshev polynomials 
 \citep[see, e.g.,][]{bros:etal:96}, sums of log-Normal functions \citep{warr:etal:13}, regularized inversion using a Lagrangian multiplier on an $L_2$ norm \citep{hann:kont:12}, pragmatic thermal-response-basis deconvolution \citep{plow:etal:13}, and an MCMC method that carries out variable smoothing across temperature space based on the widths of the emissivity curves \citep{kash:drak:98}. A number of studies have attempted to evaluate and compare the methods  \citep[see, e.g.,][]{warr:broo:09, hann:kont:12, plow:etal:13, test:etal:12}, but there is no consensus on which reconstructions are most useful or on the systematic errors or biases inherent in each. Nonetheless, the various methods exhibit  certain relative advantages and disadvantages. For example, the inversion methods are generally extremely fast, but cannot guarantee positivity of the solutions, nor do they generate uncertainty estimates. The forward-fitting methods are slower, but are adequately fast in practical situations. Although they usually produce uncertainty estimates via parametric bootstrap of the data, these methods are strongly constrained by the adopted model for regularization. The MCMC based method is the slowest, and generally cannot be applied as a black box because of the nature of MCMC,
\edit{making it difficult to apply on a pixel-by-pixel basis.}
However, it uses the most physically justifiable regularization, and generates meaningful uncertainty estimates on all parameters.

\subsection{Image segmentation}
\label{sec:imgseg}






We aim to avoid the \edit{problematic} choice of regularization \edit{and/or computational expense associated with pixel-by-pixel DEM reconstruction.} Instead we focus on identifying solar regions with similar {\it normalized} DEMs without (at least initially) specifying what those DEMs are. Such regions, if they exist, are quite interesting from a scientific perspective. Differing normalizations among otherwise similar DEMs can largely be attributed to differing volumes occupied.\footnote{\edit{This is because we expect higher electron densities
to correspond to higher temperatures, and it is unlikely that 
the electron density
and volume will conspire to produce the same 
normalized
DEM}.} Thus, regions with similar normalized DEMs are likely to have similar underlying thermal structures, even if they are spatially separated. Identifying these regions can be viewed as data preprocessing in that the clusters of pixels may be subject to secondary analyses that aim to explore their thermal properties and/or their evolution. 
\edit{This might involve fitting the DEM via (\ref{eq:lin-inv}) in each of $\sim$10 regions rather than on in each of $\sim$10$^7$ pixels. This reduces the computational expense by a factor of $\sim$10$^6$, enabling more principled regularization in each fit.}

Identifying these regions requires us to segment the  solar image. 
In particular, we seek to group pixels together according to their (unobserved) thermal distributions, $\bth_i$, while ignoring 
the variation in the nuisance parameters $\nu_i$. \edit{For simplicity, we do so without imposing model-based spatial structure on the image.}
 Ideally, we could estimate a dissimilarity function between $\bth_i$ and $\bth_j$ in each pixel pair and group pixels with small dissimilarity.
To avoid estimating $(\bth_1,\ldots, \bth_n)$, however, we base our segmentation on a surrogate dissimilarity between the observations $\by_i$ and $\by_j$, instead of the desired dissimilarity between the temperature distributions $\bth_i$ and $\bth_j$.  We aim to define a surrogate dissimilarity that leads to segmentations that are as faithful as possible to underlying clustering among the values of $\bth_i$.

\citet{stei:etal:12} also formed clusters of pixels using the observed $\by_i$ with the goal of identifying solar regions with similar underlying thermal activity. To avoid dependence on the nuisance parameter $\bnu=(\nu_1,\ldots, \nu_n)$, they formed clusters with similar normalized observed passbands, $\br_i=\by_i/||\by_i||_1$, where $||\by_i||_1 =\sum_{j=1}^b y_{ij}$. In particular, they proposed a $k$-means algorithm based on Hellinger distance instead of Euclidean distance, calling the resulting clustering algorithm H-means.  This amounts to using Hellinger distance as a surrogate dissimilarity function.  H-means is a special case of a general class of clustering algorithms based on entropy-like distances \citep{tebo:etal:06}.   The Hellinger distance is a reasonable choice both because it reduces the influence of $\bnu$ and because it accounts for heteroscedasticity via the variance stabilizing transformation for the Poisson model. \citet{stei:etal:12}, however, did not quantify the efficacy of this choice for identifying pixel clusters with similar normalized DEMs nor did they attempt to find the optimal dissimilarity function for this task. We address both of these issues \edit{within a broad class of parametrized dissimilarity measures}.

Image segmentation is one of the most well-studied problems in image processing and computer vision; among the more popular methods are those of \citet{coma:meer:02,shi:malik:00,felz:hutt:04}. Much work in this area is motivated by the challenge of segmenting images of 
scenes into distinct objects. It is extremely difficult to create appropriate criteria and efficient algorithms for this task, and \citet{estr:jeps:09} find that state-of-the-art segmentation algorithms fall short of the performance of humans segmenting images by hand \citep[][Ch.~5]{szel:10}. We do not address this task, but instead focus on segmenting solar images as a preprocessing step for later scientific analyses. As in the more traditional problem of segmenting scenes into distinct objects, segmenting multi-passband images of the Sun according to their thermal properties requires us to cluster pixels according to an appropriate similarity between pixels. However, the solar image segmentation problem is distinct when viewed as a preprocessing step because we judge methods on how well they preserve information for subsequent analyses, not on how correctly they separate distinct objects. Our approach is also different from compression/quantization frameworks \citep[e.g.,][]{gray:neuh:98} that judge methods on how well they preserve the original signal: in principle, we would not object to a segmentation that failed to preserve all features of the original signal but did preserve all information required for later statistical analyses, such as sufficient statistics.

The remainder of this article is organized into five sections. We propose to preprocess solar images by segmenting them into clusters of pixels with similar thermal activity. Section~\ref{sec:opt-preproc}  formalizes a decision theoretic approach for quantifying information loss due to this preprocessing. Section~\ref{sec:dissimilar} introduces our approach to image segmentation: we define a parameterized dissimilarity function for pairs of pixels with the aim of selecting parameters that minimize information loss. Numerical methods for selecting good parameter values for the dissimilarity function are presented in Section~\ref{sec:param} and an application to AIA data appears in Section~\ref{sec:data}.  The paper concludes in Section~\ref{sec:disc} with a discussion on the secondary analysis of our segmented images and a broader perspective on preprocessing and data reduction in the context of  
the science-driven analysis of big data.

\section{Optimal preprocessing}\label{sec:opt-preproc}

\subsection{Tunable preprocessing}

We focus on image preprocessing that involves aggregating counts across multiple pixels. In particular, we consider methods that segment images into clusters of pixels with similar features and summarize each cluster by the sum of its pixel counts in each band. This essentially imposes the assumption that $\bth_i = \bth_j$ for pixels $i$ and $j$ that are assigned to the same cluster. In this section, we address the question of how much is lost if we impose this assumption when it is not true. Of course, this can only be addressed insofar as the observed passband intensities are sensitive to differences in $\bth$.


To formally define the aggregate versions of $\by$ across clusters of pixels, 
let $\Pi(\balp)$ denote a partition of  the pixel indices, $\{1,\ldots,n\}$, into $\nclus$  clusters, $\mathcal{I}\clone, \ldots, \mathcal{I}\clnc$, where $\balp$ represents parameters of the process leading to the choice of partition.
\edit{Specifically, we let $\balp$ be the parameter of a dissimilarity measure between the observations, and the partition is the result of clustering observations according to this dissimilarity measure.} 
We can then define the aggregation of $\by$ to be
\[
	S(\by; \balp)	= \lt\{\Pi(\balp), \sum_{i \in \mathcal{I}\clone} \by_i, \ldots, \sum_{i \in \mathcal{I}\clnc} \by_i\rt\},
\]
where $S(\by; \balp)$ contains both the partition, $\Pi(\balp)$, and the sums of the multi-passband pixel counts in each of the clusters. 
Our goal is to choose  a good or even an optimal value of $\balp$ in order that the aggregated data, $S(\by; \balp)$, preserve as much information as possible for later analyses, compared to the raw counts $\by$.

\subsection{Parametric common ground}

Because we face an ill-posed inverse problem, neither the full DEM, $\nu_i \bth_i$, nor the normalized DEM, $\bth_i$, is identifiable without additional constraints or prior information. We quantify the loss of statistical information due to preprocessing in terms of a {\it sufficient identifiable parameter}. 
\edit{
Let $P_{\xi}$ denote a family of probability measures indexed by a parameter $\xi$. 
\citet{barankin:60} defines a {\it sufficient} parameter as a function $f(\xi)$ such that, for all $\xi$ and $\xi'$,  $f(\xi) = f(\xi')$ implies that $P_{\xi} = P_{\xi'}$ \citep[see also][]{picci:77,dawid:79}. An {\it identifiable} parameter, on the other hand, is a function $f(\xi)$ such that, for all $\xi$ and $\xi'$, $P_{\xi} = P_{\xi'}$ implies that $f(\xi) = f(\xi')$.  
If $f(\xi)$ is a sufficient \edit{and} identifiable parameter, and the density of the data given $\xi$ is $p(y \mid \xi)$,} then the likelihood can be expressed as $p(y \mid \xi) = p(y \mid f(\xi))$. Intuitively, sufficiency ensures that the parameter $f(\xi)$ is rich enough to capture the likelihood, and identifiability ensures that $f(\xi)$ is not too rich for the likelihood to be informative about \edit{$f(\xi)$}. 

In ill-posed inverse problems with no general consensus as to the choice of prior distribution or regularization function, a sufficient identifiable parameter can provide common ground. Such is the case for solar DEM analysis. A sufficient identifiable parameter offers an estimand that all researchers (who agree on the likelihood) can agree on as a valid inferential target, even if the researchers disagree about how best to address the ill-posedness. For this reason, we target estimators of sufficient identifiable parameters when quantifying the loss of statistical information due to preprocessing.  More generally, if there are multiple nested statistical models under consideration, we should use a sufficient identifiable parameter under the largest of these models  to measure the loss of statistical information for the widest range of possible downstream analyses. 

In model (\ref{eq:lin-inv}), an obvious sufficient identifiable parameter is  $\blam_i = \nu_i \mathbf{A R} \bth_i$, the vector of expected counts in each passband. Unfortunately, $\blam_i$ is a function of the nuisance parameter, $\nu_i$. This is undesirable: if we tailor preprocessing to best preserve estimation of ${\bs \lambda}_i$, we may (unintentionally) sacrifice the quality of estimation of the parameter of interest, $\bth_i$. In other words, we may sacrifice information that aides estimation of parameters we care about because we are tuning the conservation of information toward better estimating a nuisance parameter.  To avoid this, we decompose a sufficient identifiable parameter into a nuisance-dependent component and a nuisance-free component. We then quantify information loss due to preprocessing in terms of the quality of estimation of the nuisance-free component. 

By focusing on the nuisance-free component, we sacrifice sufficiency for the full likelihood but may maintain partial sufficiency.  In particular, consider decomposing $\blam_i$ into the nuisance-dependent component, $\sum_{j=1}^b \lambda_{ij}$, 
and the nuisance-free component,
\begin{equation}
{\bphi}_i = \blam_i / \sum_{j=1}^b \lambda_{ij} = \mathbf{A R} \bth_i / \sum_{j=1}^b \tau_j \mathbf{R}_j^{\top} \bth_i,	
\end{equation}
where $\mathbf{R}_j^{\top}$ denotes the $j$th row of $\mathbf{R}$. The likelihood for pixel $i$ factors as
\[
	p(\by_i \mid \blam_i) = p(\by_i \mid y_{i+}, \bphi_i) \ p(y_{i+} \mid \textstyle{\sum_{j=1}^b} \lambda_{ij}),
\]
where $y_{i+} = ||\by_i||_1$, $p(\by_i \mid y_{i+}, \bphi_i)$ is a multinomial distribution with size $y_{i+}$ and probability vector $\bphi_i$, and $p(y_{i+} \mid \sum_{j=1}^b \lambda_{ij})$ is a Poisson distribution with mean $\sum_{j=1}^b \lambda_{ij}$. Although the nuisance-free parameter $\bphi_i$ is not sufficient for the full likelihood  of $\by_i$, it is a sufficient identifiable parameter of the conditional distribution of $\by_i$ given $||\by_i||_1$. 


\subsection{Quantifying the loss of statistical information}\label{sub:Bayes-est}

We propose a Bayesian decision theoretic approach to choosing the tuning parameter, $\balp$, used for image segmentation. In particular, we consider minimizing the {\it expected} loss due to preprocessing. Consider an estimator $\bhatphi_i(S(\by; \balp))$ for ${\bphi}_i$, based on the aggregate data $S(\by; \balp)$. Given a loss function, $L({\bphi},\hat{\bphi})$, let $R({\bphi},\bhatphi) = E\{L({\bphi},\bhatphi) \mid \bphi\}$ be the risk, with the expectation taken with respect to $\by$. Given a prior distribution $p(\bnu, \bth)$, the optimal choice of $\balp$ is simply the Bayes estimator 
\begin{equation}
	\balp^* = \argmin_{\balp} R_B(\balp) \ \hbox{ where } \ R_B(\balp) = E\lt\{\frac{1}{n}\sum_{i=1}^n R\lt({\bphi}_i,\bhatphi_i(S(\by; \balp))\rt)\rt\},
	\label{eq:bayesrisk}
\end{equation}
is the Bayes risk and the expectation in (\ref{eq:bayesrisk}) is taken with respect to $p(\bnu, \bth)$. We use the notation $\balp^*$ instead of $\hat\balp$ to emphasize that $\balp^*$ is simply an optimal choice of tuning parameter for a preprocessing step, not an estimator of a scientifically meaningful parameter.

In practice, preprocessing often involves somewhat ad hoc or heuristic algorithms. In image segmentation, the clustering of pixels typically comes down to an iterative algorithm such as $k$-means. In these cases, it is hopeless to attempt to obtain $\balp^*$ analytically. Instead, we approximate the Bayes risk via simulation from a scientifically motivated prior distribution $p(\bnu, \bth)$, 
\edit{and find the optimal $\balp^*$ via a grid search.}
The computational expense is somewhat mitigated by the fact that we limit the size of the simulated datasets for this tuning stage. 

\section{Tunable dissimilarities}
\label{sec:dissimilar}

\subsection{Pairwise dissimilarity functions}
\label{sec:diss-funct}

Image segmentation is fundamentally a clustering problem. It involves clustering pixels into groups corresponding to distinct---but not necessarily contiguous---image regions. One strategy is to define a  numerical dissimilarity between each pair of pixels and then to optimize the partition of pixels into clusters such that within-cluster dissimilarities are small. This is typically accomplished via a function which returns the dissimilarity or simply the distance between any pair of pixels. For example, a common approach is to use a weighted sum of the spatial distance and the distance between pixel-specific covariates such as intensity, color, or texture \citep[e.g.,][]{shi:malik:00}.  As discussed in Section~\ref{sec:imgseg}, the spatial distance 
\edit{is not of primary interest in our case and may not be pertinent, so that}
dissimilarity is based solely on differences in the observed photon counts in each passband. 

The Minkowski 
distance, 
\[
\dmin(\by_i, \by_j) = || \by_i - \by_j||_p, \ \hbox{ \ where \ } \ ||\bx||_p = \lt\{\sum_{l=1}^b |x_l|^p\rt\}^{1/p},
\]
offers a general class of dissimilarity functions, including the special cases of Euclidean distance ($p=2$), Manhattan distance ($p=1$), and Chebyshev distance (in the limit as $p\to\infty$). 
The Euclidean distance in particular is ubiquitous in image segmentation and more general clustering problems. In our setting, however, any Minkowski distance between vectors of raw counts, $\by_i$ and $\by_j$, is strongly influenced by the nuisance parameters, $\nu_i$ and $\nu_j$. As illustrated numerically in the case of Euclidean distance in Section~\ref{sec:recover}, this can overwhelm comparisons between the  corresponding 
normalized DEMs,
$\bth_i$ and $\bth_j$, which are of primary interest. 
The total count in each pixel, $ ||\by_i||_1$, is a sufficient statistic for the nuisance parameters $\nu_i$, and thus we prefer dissimilarity functions that depend on the pixel counts only through their $L_p$ normalized values, $\by_i /||\by_i||_p$. While $\by_i /||\by_i||_1$ is an obvious choice,  we shall see that $L_1$ and $L_2$ normalizations are related in such a way that we lose nothing by focusing first on the $L_2$ normalization (see (\ref{eq:cos-hell-general}) below).


A choice of dissimilarity function that achieves this is the cosine dissimilarity,
\begin{equation}
\dcos(\by_i, \by_j) = \frac{1}{2} \ \deuc^2 \left(\frac{\by_i}{||\by_i||_2}, \frac{\by_j}{||\by_j||_2}\right) 
	= 1 - \frac{\by_i^{\top}\by_j}{||\by_i||_2 \ ||\by_j||_2};
\label{eq:cos-euc}
\end{equation}
this is one minus the cosine of the angle between the vectors $\by_i$ and $\by_j$. Cosine dissimilarity is widely used in text mining applications in which documents are compared by word frequencies with the goal of identifying documents with similar word distributions while ignoring document length \citep[e.g.,][]{hoth:etal:05,huan:08}.

\subsection{Parameterizing dissimilarity functions}
\label{sec:param-diss-funct}

To formulate a more flexible class of dissimilarity functions, we consider parametric dissimilarities $D(\by_i, \by_j; \balp)$, where $\balp$ is a vector of parameters that specify the dissimilarity.  In particular, we introduce $\balp$ through a transformation of $\by_i$. We limit our attention to transformations having the same dimension as $\by_i$, which can be denoted $\bT(\by; \balp) = \lt(T_1(\by;\balp), \ldots, T_b(\by;\balp)\rt)$. The parameterized dissimilarity is simply a familiar distance or dissimilarity between transformed observations, i.e., 
\begin{equation}
	D(\by_i, \by_j; \balp) = d\lt(\bT(\by_i; \balp), \bT(\by_j; \balp)\rt),
	\label{eq:dist-trans}
\end{equation}
where $d(\cdot,\cdot)$ could be, for example, Euclidean distance or cosine dissimilarity.

There is a large literature on learning distance metrics. One common approach \citep[see, for example,][]{xing:etal:02,shal:etal:04,weinberger:etal:05,davis:etal:07,jain:etal:12} is to assume that the parameter $\balp$ is a matrix, say ${\bs M}$, and to find ${\bs M}$ such that the Euclidean distances between the linearly transformed observations $||{\bs M} \by_i - {\bs M}\by_j||_2$ satisfy an optimality criterion, such as increasing distances between pairs of training observations labeled dissimilar and decreasing distances between pairs labeled as similar. Motivated by the problem of facial recognition, \citet{nguy:etal:11} consider learning cosine dissimilarities instead of Euclidean distances, but also focus on linear transformations.


Instead of linear transformations, we consider the family of power transformations,
\begin{equation}
	T_j(\by_i; \balp) = (y_{ij} + \gamma)^{\beta}, \quad \hbox{for} \quad j = 1,\ldots,b, \label{eq:T}
\end{equation}
where $\beta > 0$, $\gamma \geq 0$, and $\balp = (\beta, \gamma)$. 
Special cases of (\ref{eq:T}) include the variance stabilizing transformation for a Poisson random variable  $(\beta = 1/2, \gamma=0)$ and one half of the Anscombe (\citeyear{ansc:48}) transform $(\beta = 1/2, \gamma= 3/8)$. Following (\ref{eq:dist-trans}),  we explore the use of cosine dissimilarities between observations transformed according to (\ref{eq:T}), i.e., we use
\begin{equation}
	\Dcos(\by_i, \by_j; \balp) = d_{\cos}\Big(\bT(\by_i; \balp), \bT(\by_j; \balp) \Big). \label{eq:cos-transformed}
\end{equation}
The Hellinger distance between the two probability vectors, $\br_i=\by_i /||\by_i||_1$ and $\br_j=\by_j /||\by_j||_1$, is a special case because, if $\balp =(1/2, 0)$, then
\begin{equation}
\Dcos (\br_i, \br_j; \balp)
 = \frac{1}{2} \deuc^2 \Big(\bT(\br_i; \balp), \bT(\br_j; \balp) \Big)
 = \frac{1}{2}\sum_{l=1}^b \Big(\sqrt{r_{il}} - \sqrt{r_{jl}} \Big)^2
 = \dhell^2(\br_i, \br_j)
\end{equation}
 where the first equality follows from (\ref{eq:cos-euc}). More generally,
\begin{equation}
	\Dcos\Big(\by_i, \by_j; (\beta, \gamma) \Big) = 
	\dhell^2\lt(\frac{\bT(\by_i; (2\beta, \gamma))}{||\bT(\by_i; (2\beta, \gamma))||_1}, \frac{\bT(\by_j; (2\beta, \gamma))}{||\bT(\by_j; (2\beta, \gamma))||_1}\rt); \label{eq:cos-hell-general}
\end{equation}
see Appendix~\ref{sec:proof} for details.
This final expression means that the class of dissimilarities in (\ref{eq:cos-transformed}) encompasses not only the ($L_2$-normalized) Euclidean distance or cosine dissimilarity, but also the ($L_1$-normalized) Hellinger distance. 

Stein {\it et al.}'s (\citeyear{stei:etal:12}) use of  $\dhell^2(\br_i, \br_j)$ to segment solar images can be justified by a desire to reduce the influence of the nuisance parameters, $\nu_i$, through the use of the cosine dissimilarity and to stabilize the variance of observations with different total count via the square root transformation, specified here by $\balp = (1/2,0)$.  Together, these considerations provide a principled justification for the use of Hellinger distance  in \citet{stei:etal:12}. In this paper, we consider the possibility that other choices of $\balp$ may lead to better performance. Section~\ref{sec:param} develops a numerical framework for computing good values of $\balp$. 

\subsection{Clustering \edit{and Numerical Issues}}
\label{sec:clust}

Once we have identified a dissimilarity function, the next step is to partition the pixels into clusters that (approximately) minimize the within-cluster dissimilarities. Given the relationship between the cosine dissimilarity and Euclidean distance given in (\ref{eq:cos-euc}), this can easily be accomplished using the $k$-means algorithm. In particular, for any choice of $\balp$ the parametrized dissimilarity function in (\ref{eq:cos-transformed}) can be written
\begin{equation}
	\Dcos(\by_i, \by_j; \balp) 
	= \frac{1}{2}d^2_2\lt(\frac{\bT(\by_i; \balp)}{||\bT(\by_i; \balp)||_2}, \frac{\bT(\by_j; \balp)}{||\bT(\by_j; \balp)||_2 } \rt).
\label{eq:param-dissim}
\end{equation}
Applying the $k$-means algorithm to the transformed observations $\bT(\by_i; \balp) /||\bT(\by_i; \balp)||_2$ efficiently partitions pixels by minimizing the sum of squared Euclidean distances between the transformed observations and their nearest cluster centroids.

\edit{Unfortunately, 
the final partition produced by $k$-means can be sensitive to the initial locations of cluster centroids. To numerically mitigate this, we  use the $k$-means++ algorithm of \citet{arthur2007k} to choose the initial centroids. This is a randomized algorithm that encourages dispersed initial centroids. We rerun $k$-means++ with five different random seeds and choose the resulting partition (from the five) that yields the minimum within-cluster sum of squared distances. 

Using $k$-means requires us to choose $\nclus$, the number of clusters. 
Although there are many available methods that accomplish this \citep[e.g.,][]{mill:coop:85,tibs:etal:01}, we do not view this choice as a statistical problem, at least in this setting.  This is because there is not some number, $\nclus < n$, of distinct DEM types and so there is not a true number of clusters that we are trying to estimate. Instead, we view the choice of $\nclus$ as governing the degree of image compression: fewer clusters correspond to higher image compression. Therefore, we anticipate that in practice the choice of $\nclus$ will be guided by considerations 
that are not just statistical (e.g., bandwidth and storage limitations), and thus we do not propose a specific method here for choosing $\nclus$.} 


\edit{In practice, we recommend spatially smoothing the images before applying $k$-means. From a scientific point of view, this allows us to focus on solar structures of interest (e.g., loops) and to suppress smaller features. Smoothing also stabilizes the cosine dissimilarities which can be noisy when the denominators in (\ref{eq:param-dissim}) are small. Smoothing effectively takes advantage of spatial similarities to yield better estimates of intensities in regions with few counts, reducing the noise in the ratios in (\ref{eq:param-dissim}).}

\section{Numerical selection of dissimilarity functions}\label{sec:param}


\subsection{Simulation design}\label{sub:sim}

In practice we must determine what value of $\balp$ to use in the definition of the dissimilarity function. We propose to do so numerically, by conducting a simulation study and choosing the optimal value of $\balp$ (e.g., the value that minimizes the Bayes risk) in the simulation. If the simulation is designed to mimic the scientific setting of interest, the optimal value of $\balp$ under the simulation can be expected to behave well in the actual scientific data analysis. 
\edit{To ensure this, we rely on subjective, informative prior distributions (see Table~\ref{tbl:prior}) that are based on our experience with the range of potential shapes of DEMs. For example, solar DEMs often have one or two peaks between approximately $10^6$ and $10^7$ K; see \citet{kash:drak:98} for numerous examples. To allow for the possibility of unexpected DEM shapes, we choose hyperparameters that yield prior distributions that are somewhat more diffuse than our subjective prior expectations.}

We conduct a pair of simulation studies in which there are two 
distinct clusters among the $(\bth_1,\ldots,\bth_n)$. We denote the centroids of the two clusters by ${\bs \vartheta}\clone$ and ${\bs \vartheta}\cltwo$ and write 
$$\bth_i \simeq \
\begin{cases}
{\bs \vartheta}\clone &\mbox{if pixel $i$ belongs to cluster 1}  \\
{\bs \vartheta}\cltwo &\mbox{if pixel $i$ belongs to cluster 2} 
\end{cases}
$$
and $\bth_{i_1} \simeq \bth_{i_2}$ if pixels $i_1$ and $i_2$ are in the same cluster. This induces a partition of the pixels and we let $\mathcal{I}\clj = \{i : \bth_i \simeq {\bs \vartheta}\clj\}$ for $c=1,2$. (For simplicity in this section, we assume that the number of clusters is \edit{well defined and} known.) The nuisance parameters, $(\nu_1, \ldots, \nu_n)$, on the other hand, may take on values unrelated to this partition. This means that while $i_1, i_2\in\mathcal{I}\clj$ implies that $\bth_{i_1}\simeq\bth_{i_2}$, it does not imply any relationship between $\nu_{i_1}$ and $\nu_{i_2}$. Scientifically, this corresponds to assuming that there are a limited number of distinct thermal profiles (i.e., temperature distributions) but that the amount of plasma along any line of sight varies independently of the thermal profile.



\begin{table}[t]
\centering \small
\caption{The prior distribution for $\bmu_i = \nu_i\bth_i$ parameterized in terms of $\bs\psi_i =(\pi_i, m_{i1}, m_{i2}, \sigma_{i1},\sigma_{i2}, \nu_i)$ as in (\ref{eq:DEM-mixture}).
\label{tbl:prior}}
\begin{tabular}{l l}
&\\
\toprule
\multicolumn{1}{c}{Parameter} & \multicolumn{1}{c}{Explanation} \\
\hline
$\pi_i \sim \text{\sc Beta}(3.5, 3.5)$, & $\Pr(0.1 < \pi_i < 0.9) = 99\%$\\
\\
$m_{i1}, m_{i2} \iid \text{\sc Uniform}[5.5, 8.0]$ & Uniform over a plausible range of $\log_{10}$(temperature) \\
\\
$\log\sigma_{i1}, \log\sigma_{i2} \iid {\rm N}(-1.8, 0.7^2)$ & 
$\E(\sigma_{ij})\approx 0.2$ and $\Pr(0.025 < \sigma_{ij} < 1.0) = 99\%$\\
\\
$\log_{10} \nu_i\sim\text{\sc Uniform}[26.5, 28]$, &  For a uniform DEM, this prior implies that\\ 
& $\E\lt(\sum_{j=1}^b y_{ij}\rt)$ ranges from approximately $50^\dagger$ to $1500$\\
\bottomrule
\multicolumn{2}{l}{\footnotesize $^\dagger$Although $50$ is relatively high  for the minimum expected count under a uniform DEM, the }\\
\multicolumn{2}{l}{\footnotesize \ \; expected count can be much lower for a non-uniform DEM.}
\end{tabular}
\end{table}

\begin{figure}[t]
\includegraphics[width=1.00\textwidth]{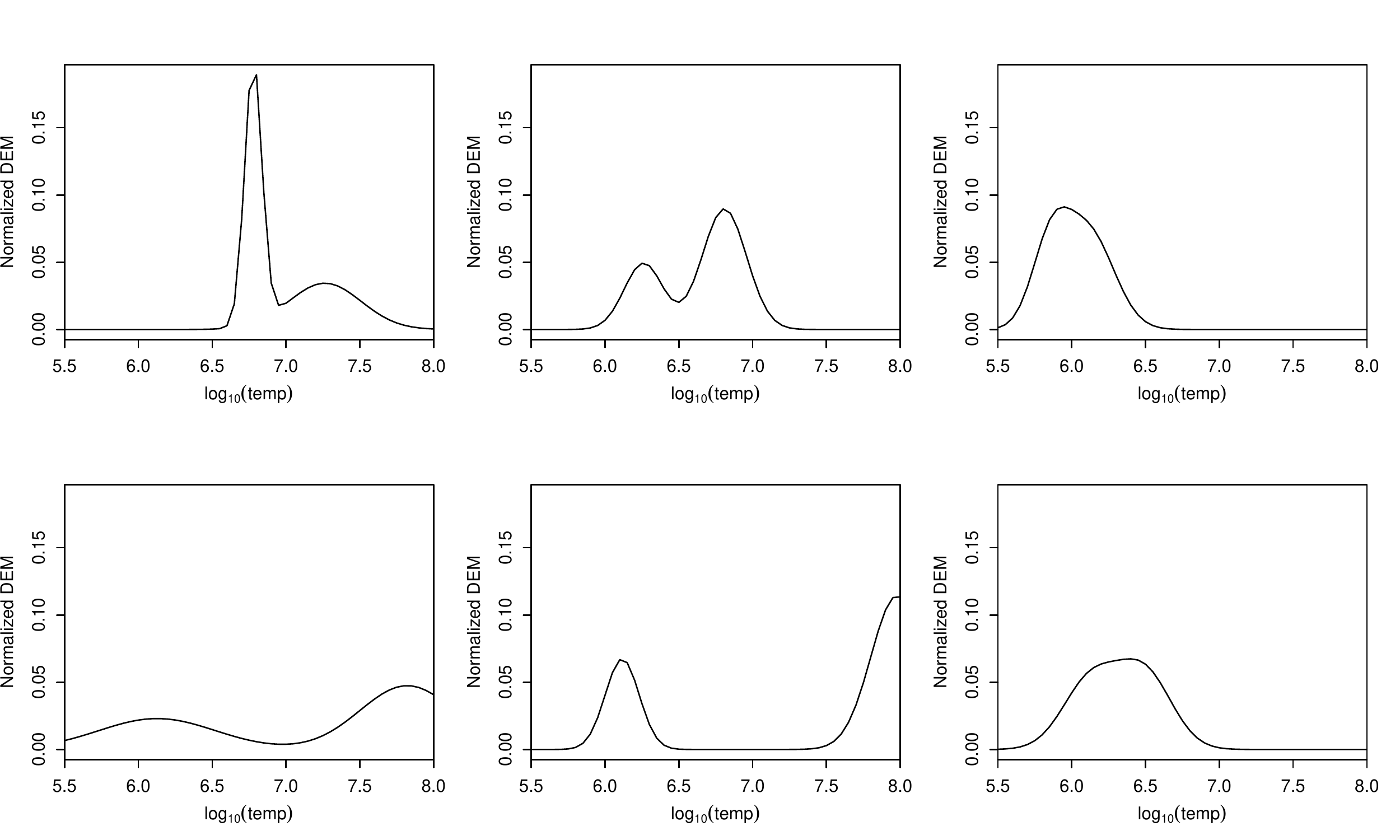}
\caption{A sample of normalized DEMs parameterized as in (\ref{eq:DEM-mixture}) and simulated with the prior distributions described in Table~\ref{tbl:prior}. \label{fig:DEM-mixture}}
\end{figure}

To mimic the scientific expectation of the variability among the possible shapes of the solar DEM, we suppose that each
$\bmu_i$ is constructed as a discretized version of a truncated mixture of two Gaussian density functions. In particular, we let $\mu_{ij} = \nu_i \theta_{ij}$, where
\begin{equation}
	\theta_{ij} = \frac{1}{A} \lt\{\pi_i \frac{1}{\sigma_{i1}}\phi\lt(\frac{\log_{10} T_j - m_{i1}}{\sigma_{i1}}\rt) + (1-\pi_i) \frac{1}{\sigma_{i2}}\phi\lt(\frac{\log_{10} T_j - m_{i2}}{\sigma_{i2}}\rt)\rt\},\label{eq:DEM-mixture}	
\end{equation}
$\log_{10} T_j = 5.5, 5.55, \ldots, 8.0$ is a grid of values for the coronal temperature, $\phi(\cdot)$ is the standard normal density function, $\pi_i$ is the relative size of the first Gaussian distribution, $m_{i1}$ and $m_{i2}$ are the means of the component Gaussian distributions, $\sigma_{i1}$ and $\sigma_{i2}$ are the standard deviations, and 
\[
	A = \sum_{j=1}^{\ntemp} 
	\lt\{\pi_i \frac{1}{\sigma_{i1}}\phi\lt(\frac{\log_{10} T_j - m_{i1}}{\sigma_{i1}}\rt) + (1-\pi_i) \frac{1}{\sigma_{i2}}\phi\lt(\frac{\log_{10} T_j - m_{i2}}{\sigma_{i2}}\rt)\rt\}.
\]
This specification of $\bmu_i$ depends on the unknown parameter $\bs\psi_i =(\pi_i, m_{i1}, m_{i2}, \sigma_{i1},\sigma_{i2}, \nu_i)$. We can set a prior distribution on $\bmu_i$ (or equivalently on  $\bth_i$ and $\nu_i$) by specifying a prior distribution on $\bs\psi_i$. For consistency of notation, we use $p(\bth_i, \nu_i)$ to denote our choice for the prior on $\bs\psi_i$, which is given in Table~\ref{tbl:prior}.  
A sample of normalized DEMs, $\bth_i$, parameterized as in (\ref{eq:DEM-mixture}) and simulated via $p(\bth_i, \nu_i)$, appears in Figure~\ref{fig:DEM-mixture}.

To illustrate how simulation studies can be used to select $\balp$, we consider images with $n=100$ pixels that can be partitioned into two clusters, each with 50 pixels. We conduct two simulations. In {\sc Simulation I}, there is no variability among the normalized DEMs within each cluster. That is, $i_1, i_2\in\mathcal{I}\clj$ implies that $\bth_{i_1}=\bth_{i_2}$. In {\sc Simulation II}, on the other hand, there are two clusters, each with similar, but not necessarily equal, normalized DEMs, That is, $i_1, i_2\in\mathcal{I}\clj$ implies that $\bth_{i_1}\simeq\bth_{i_2}$. Thus in {\sc Simulation I} the clusters are {\it noiseless} and in {\sc Simulation II} they are {\it noisy}.

\begin{figure}[t]
	\begin{minipage}[b]{0.3\linewidth}
	\centering
	(a)\\
	\includegraphics[width=\textwidth]{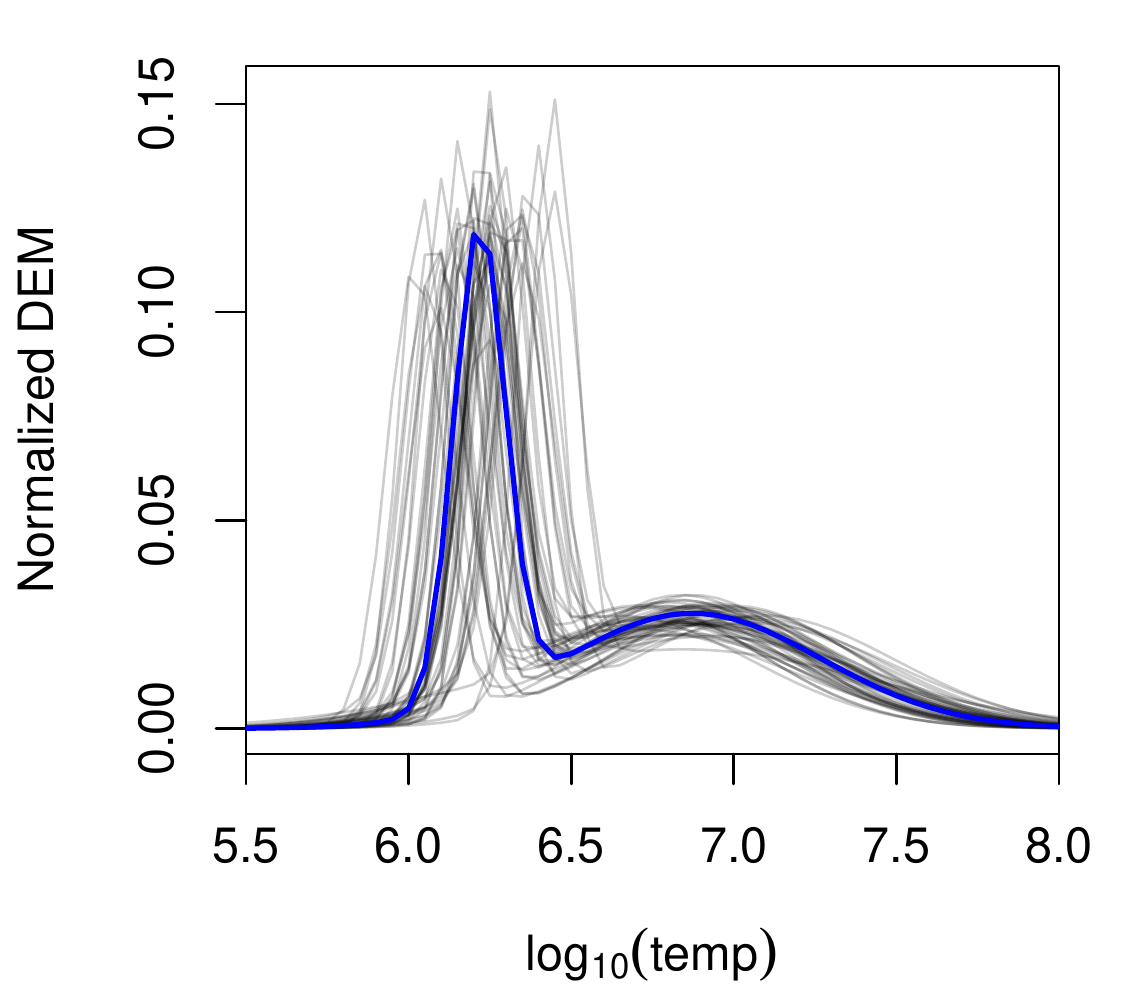}\\

	(b)\\
	\includegraphics[width=\textwidth]{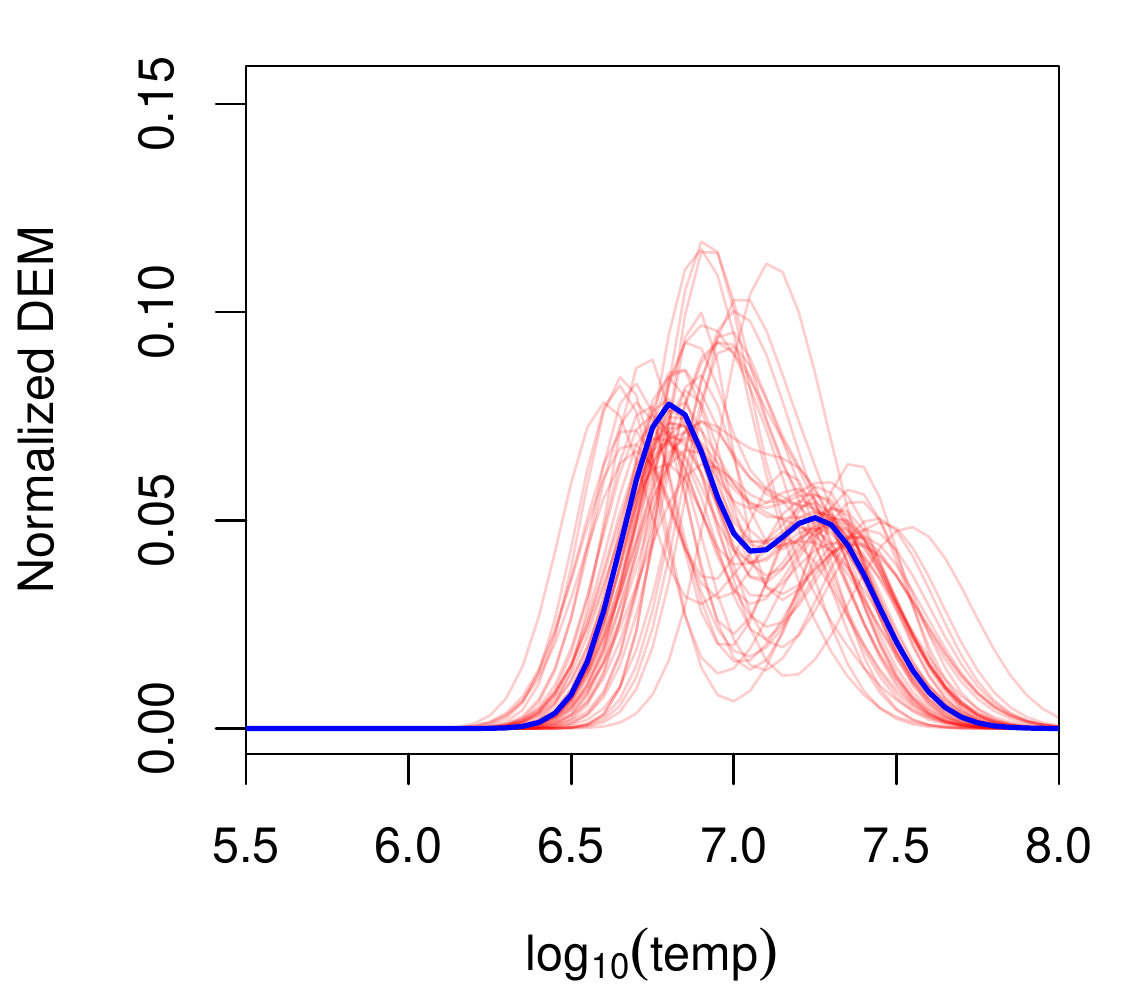}
	\end{minipage}
	\begin{minipage}[b]{0.63\linewidth}
	\centering
	(c)\\
	\includegraphics[width=.95\textwidth]{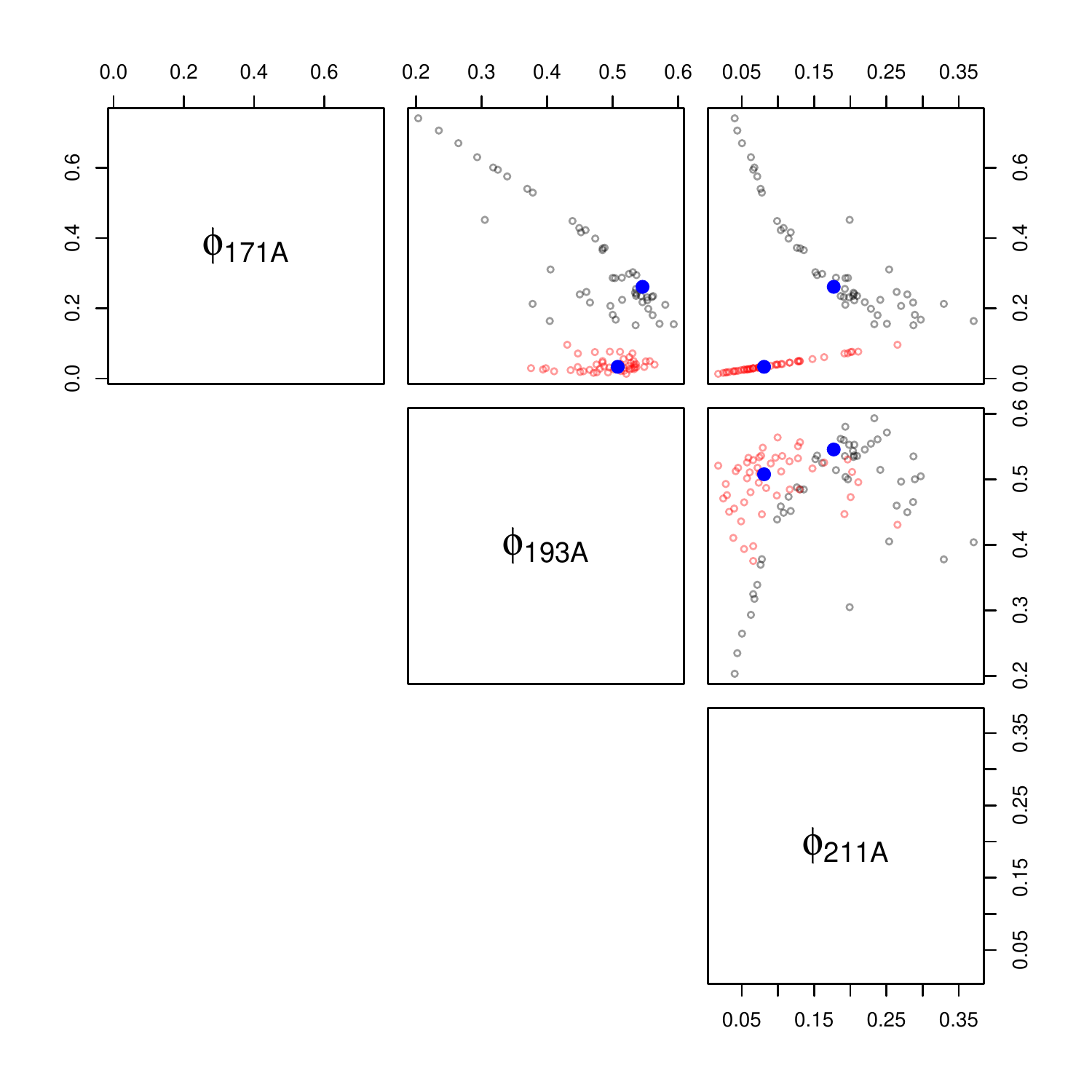}
	\end{minipage}
\caption{A comparison of the noiseless and noisy clustering of the normalized DEMs used in {\sc Simulations~I} and II, respectively. The normalized DEMs plotted in blue in panels (a) and (b) represent $\bs\vartheta_1$ and $\bs\vartheta_2$ while those plotted in black and red represent the corresponding samples of $\bth_i$ used in one of the images sampled under {\sc Simulation~II}. Panel (c) plots the resulting normalized expected photon counts, $\bphi$, in three of the six passbands, using the same color coding. \label{fig:DEM-clusters}}
\end{figure}

In each simulation, we generate $5000$ such images. Specifically, letting $\ell=1, \ldots, 5000$ index the simulated images, we first sample $\bs\vartheta\clone\iter$ and $\bs\vartheta\cltwo\iter$ from $p(\bth_i)$ as described in Table~\ref{tbl:prior} and in (\ref{eq:DEM-mixture}). In {\sc Simulation~I} we noiselessly set 
$$
\bth_i\iter = 
\begin{cases}
\bs\vartheta\clone\iter &\mbox{ for } i=1,\ldots, 50,  \\ \\
\bs\vartheta\cltwo\iter &\mbox{ for } i=51,\ldots, 100.
\end{cases}
$$
In {\sc Simulation II} we set $\pi\clj\iter, m_{(c)1}\iter,  m_{(c)2}\iter, \sigma_{(c)1}\iter$, and $\sigma_{(c)2}\iter$ to the values corresponding to $\bs\vartheta\iter\clj$ for the two cluster centroids, $c=1,2$, and noisily sample
 	\begin{align*}
 		\pi_i\iter &= 
 		\begin{cases}
 		\pi\clone\iter, &\mbox{ for } i = 1,\ldots, 50, \\ \\
 		\pi\iter\cltwo, &\mbox{ for } i = 51,\ldots, 100,
 		\end{cases} \\  \\
 		m_{ij}\iter &\sim \begin{cases}
 		{\rm N}(m_{(1)j}\iter, 0.1^2), &\mbox{ for }  i = 1, \ldots, 50 \mbox{ and } j = 1,2, \\ \\
 		{\rm N}(m_{(2)j}\iter, 0.1^2), &\mbox{ for }  i = 51, \ldots, 100 \mbox{ and } j=1,2,
 		\end{cases}
 		\\ \\
 		\log\sigma_{ij}\iter &\sim
 		\begin{cases}
 		{\rm N}(\log\sigma_{(1)j}\iter, 0.1^2), &\mbox{ for } i = 1, \ldots, 50 \mbox{ and } j = 1,2, \\ \\
 		{\rm N}(\log\sigma_{(2)j}\iter, 0.1^2), &\mbox{ for } i = 51, \ldots, 100 \mbox{ and } j=1,2.
 		\end{cases}
 	\end{align*}
An example of two noisy DEM clusters sampled in this way appears in Figure~\ref{fig:DEM-clusters}, along with a comparison with the noiseless clusters used in {\sc Simulation I}.  Finally, in both simulations, for each pixel, $i$, we sample $\nu_i\iter\iid p(\nu)$, compile $\bmu_i\iter$,
and simulate the six passband counts as independent Poisson random variables with means given in (\ref{eq:lin-inv}) with $\bmu_i$ replaced by $\bmu\iter_i$.  We denote the simulated images by $\by\iter$ for $\ell=1,\ldots, 5000$. For the results reported in this paper, we simulated the six passband counts from the truncated Poisson distribution conditional on $y\iter_{i+} > 0$. In results not reported here, we repeated {\sc Simulations~I} and {\sc II} without this truncation but with the restriction that $\gamma > 0$ (so that a strictly positive pseudo count was added to every observation when computing pairwise dissimilarities); the results were qualitatively similar to those reported. 

We consider two methods to evaluate the choice of $\balp$, its ability to recover the true underlying partition and its estimated Bayes risk. Results for the two criteria under our simulation study are reported in the next two sections.

\subsection{Recovering a true partition}
\label{sec:recover}

We first use the adjusted Rand index to evaluate the choice of $\balp$ in terms of its ability to recover the true underlying partition of the pixels. The \citet{rand:71} index is computed by summing (a) the number of pairs of units that are in the same cluster in truth and in the reconstruction and (b) the number of pairs of units  that are in different clusters in truth and in the reconstruction,  and dividing by the total number of pairs of units. The adjusted Rand index \citep{hubert:arabie:85} attempts to adjust the Rand index for chance groupings. As with the Rand index, an  exact recovery of the true clustering corresponds to an adjusted Rand index of $1$.

\begin{figure}[p]
\centering
	\begin{minipage}[b]{0.46\linewidth}
	\centering
	(a)\\
	\includegraphics[width=\textwidth]{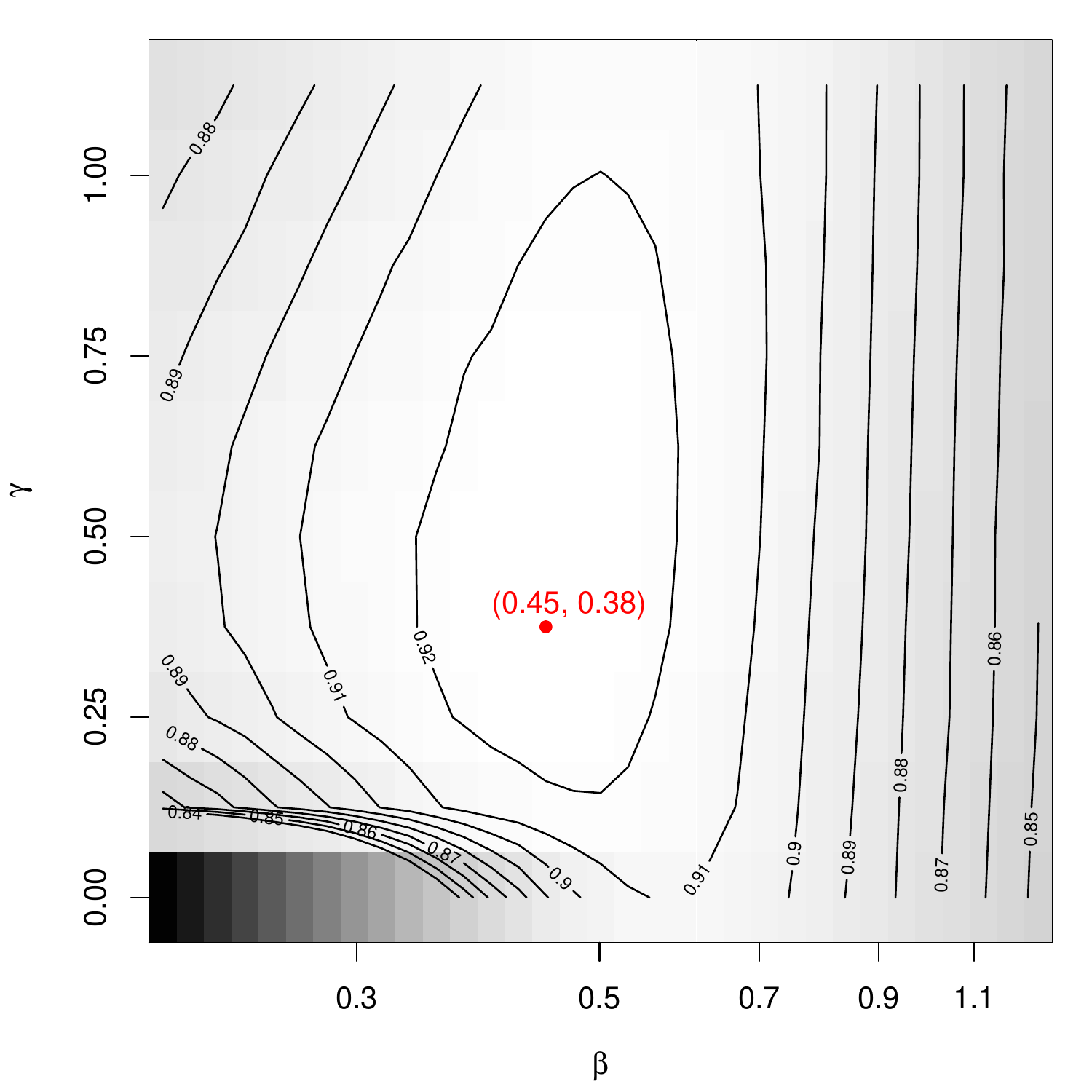}
	\end{minipage}
	\hspace{.05\linewidth}
	\begin{minipage}[b]{0.46\linewidth}
	\centering
	(b)\\
	\includegraphics[width=\textwidth]{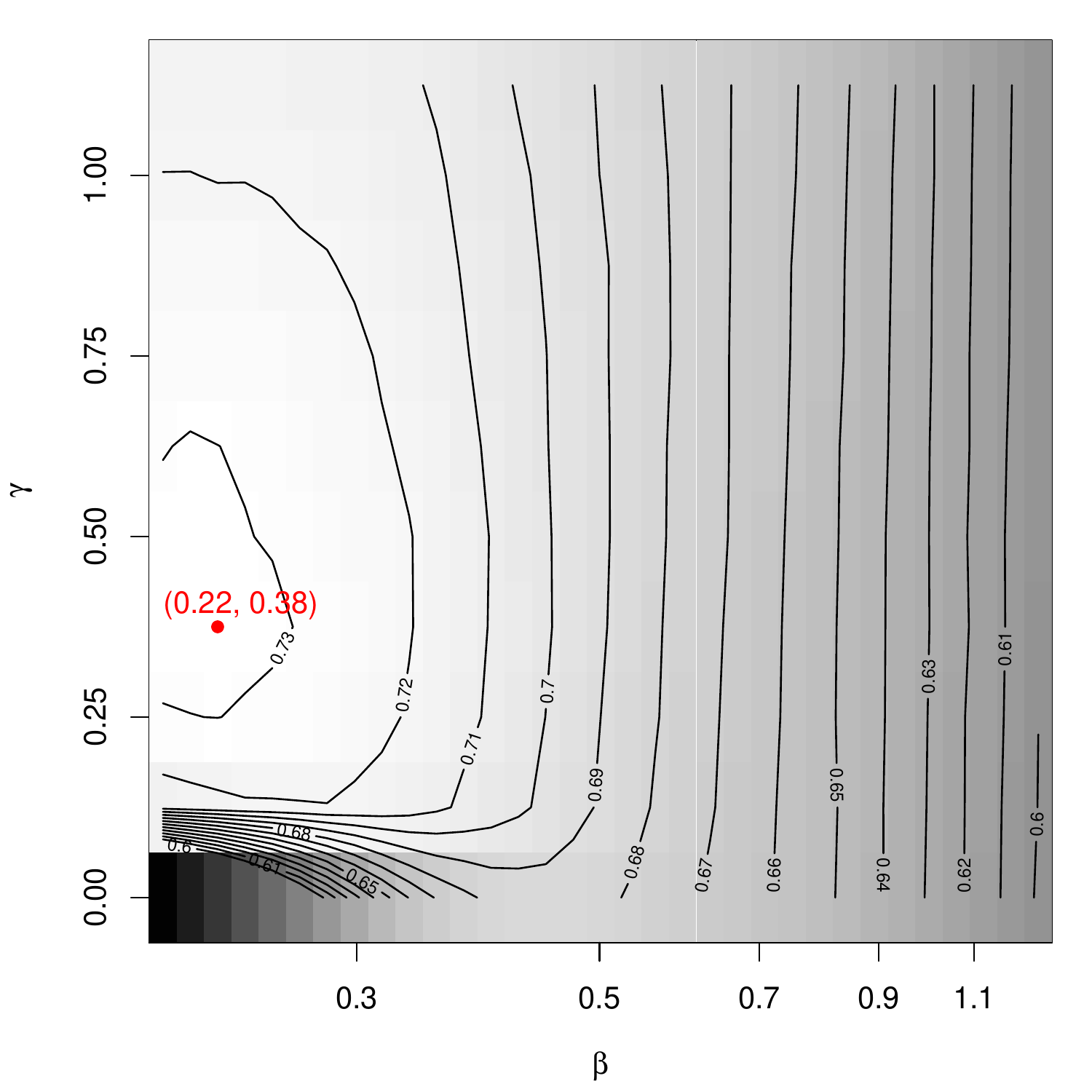}
	\end{minipage}
	\begin{minipage}[b]{0.46\linewidth}
	\centering
	(c)\\
	\includegraphics[width=\textwidth]{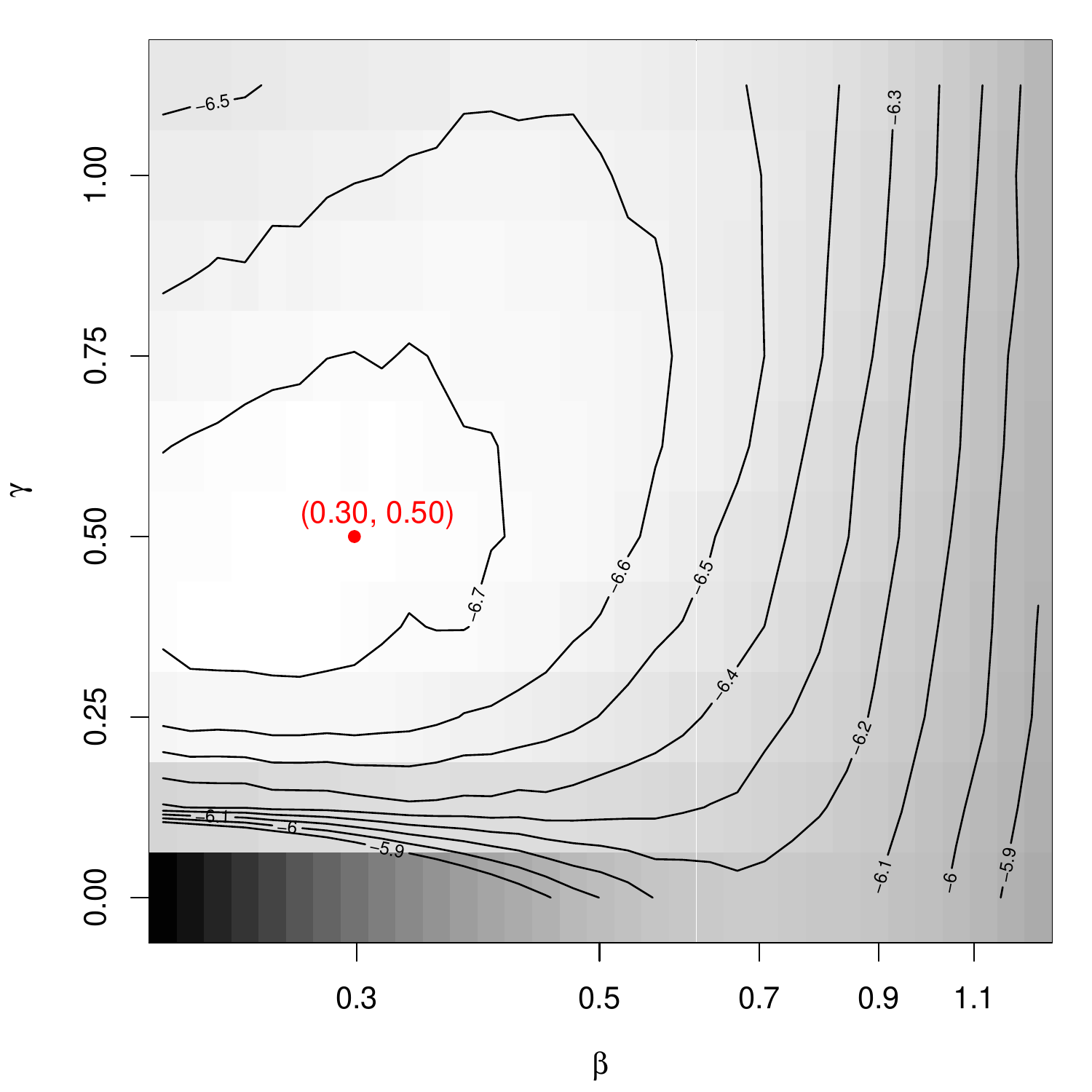}
	\end{minipage}
	\hspace{.05\linewidth}
	\begin{minipage}[b]{0.46\linewidth}
	\centering
	(d)\\
	\includegraphics[width=\textwidth]{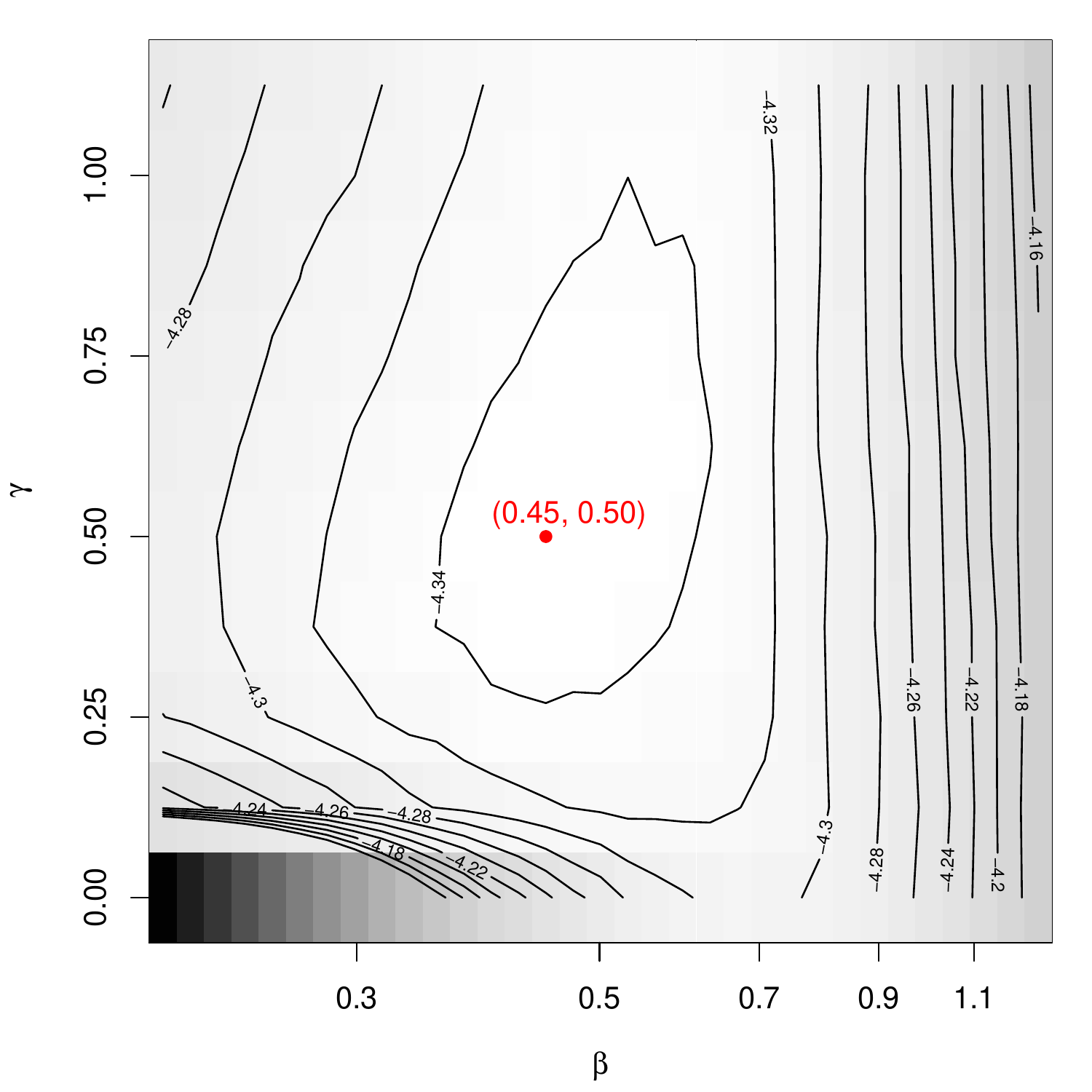}
	\end{minipage}
	\caption{Average adjusted Rand index for (a) {\sc Simulation I} and (b) {\sc Simulation II}, and log Bayes risk for (c) {\sc Simulation I} and (d) {\sc Simulation II} for a range of values of $\balp =(\beta,\gamma)$. Gray scales are chosen so that lighter shades correspond to better performance. Red points indicate the locations of maximum adjusted Rand index in (a) and (b) or minimum Bayes risk in (c) and (d). While the four comparisons do not agree precisely on the optimal value of $\balp$, they unanimously exclude the untransformed cosine dissimilarity ($\beta=1, \gamma=0$) and highlight the advantage of adding fractional pseudo counts when using the Hellinger distance (if $\beta=1/2$, $\gamma$ should be positive).} 
	\label{fig:tuning}
\end{figure}

Figure~\ref{fig:tuning}(a) plots the adjusted Rand index for {\sc Simulation I}, averaged over the 5000 simulated images described in Section~\ref{sub:sim}.  The best performance is achieved when $\beta$ is near $1/2$, but the optimal $\beta$ depends on $\gamma$: if $\gamma = 0$, then the optimal $\beta > 1/2$, whereas for $\gamma > 0$, the optimal $\beta$ is slightly less than $1/2$. Among the class of dissimilarities considered, optimal recovery of the underlying clusters is achieved at $\beta^* = 0.45$ and $\gamma^*=0.38$, with an adjusted Rand index of $0.925$.

The average adjusted Rand index for {\sc Simulation II} is shown in Figure~\ref{fig:tuning}(b). The optimal adjusted Rand index is $0.73$; this is substantially lower than for {\sc Simulation I} because it is much harder to recover the true partition when the clusters of normalized DEMs are noisy. Under the simulation design used in {\sc Simulation II}, the normalized expected photon counts, $\bphi_i$, occasionally exhibit substantial overlap between the two clusters. As in {\sc Simulation I}, the best performance is achieved when $\gamma^* = 0.38$. Here, the optimal choice of $\beta^* = 0.22$ is lower than in {\sc Simulation I}. Using the {\sc Simulation~I}-optimal value of $\balp^*=(0.45, 0,38)$ in {\sc Simulation~II}, however, reduces the adjusted Rand index by only $0.03$.

For comparison, we perform standard $k$-means clustering in both simulations. That is, we cluster the untransformed $(\beta=1, \gamma=0)$ counts using squared Euclidean distance instead of the cosine dissimilarity. This results in an average adjusted Rand index of just $0.08$ for {\sc Simulation I} and $0.06$ for {\sc Simulation II}.
The optimal $\balp$-transformed cosine dissimilarity offers a substantial improvement over this baseline. 

\subsection{Decision theoretic choice of $\balp$}

We now consider choosing an optimal $\balp$ under the decision theoretic framework of Section~\ref{sec:opt-preproc}. Here we use the quadratic loss function $L({\bphi}, \hat{\bphi})= ||\hat{\bphi} - {\bphi}||_2^2$, \edit{but in practice this could be replaced by any reasonable loss function}. We can then approximate the Bayes risk via Monte Carlo.  For example, using the $n=5000$ images simulated in Section~\ref{sub:sim}, 
\begin{equation}
	\hat R_B(\balp) = \frac{1}{5000 n}\sum_{\ell=1}^{5000}\sum_{i=1}^{n} R\lt({\bphi}_i,{\bhatphi}_i\big(S(\by\iter; \balp)\big)\rt),
	\label{eq:est-b-risk}
\end{equation}
where $\by\iter$ are sampled from their prior predictive distribution as described in Section~\ref{sub:sim} and thus average over both $p(\by \mid \bth, \bnu)$ and over $p(\bth, \bnu)$. The estimator of ${\bphi}_i$ in (\ref{eq:est-b-risk}) is given by 
\begin{equation}
	\hat{\phi}_{ij}\lt(S\big(\by\iter; \balp\big)\rt) = \frac{\sum_{l \in \mathcal{I}[i]} y\iter_{lj}}{\sum_{l \in \mathcal{I}[i]} \sum_{j'=1}^by\iter_{lj'}}, \quad j = 1, \ldots, b,
	\label{eq:phi-hat}	
\end{equation}
where $\mathcal{I}[i]$ denotes the cluster containing pixel $i$. Note that ${\bhatphi}_i$ is the maximum likelihood estimator of $\bphi_i$ under the model that assumes that $\bth_{i_1} = \bth_{i_2}$ for all $i_1, i_2 \in \mathcal{I}\clj$, for each cluster $c$. (The denominator of (\ref{eq:phi-hat}) is never zero because we simulate images conditional on $y\iter_{i+}>0$; see Section~\ref{sub:sim}.)

Figures~\ref{fig:tuning}(c) and (d) plot $\log\big(\hat{R}_B(\balp)\big)$ for a range of  $\beta$ and $\gamma$, for {\sc Simulations I} and {\sc II}, respectively. For {\sc Simulation I}, the minimum Bayes risk is achieved at $\beta^* = 0.3$ and $\gamma^*=0.5$; for {\sc Simulation II}, at $\beta^* = 0.45$ and $\gamma^* = 0.5$. Adding a positive fractional pseudo count provides a substantial benefit in {\sc Simulation I}: the optimal Bayes risk at $\gamma^* = 0.5$ is 45\% lower than the best Bayes risk that can be achieved with $\gamma = 0$. In {\sc Simulation II}, the contrast is less dramatic, with an optimal Bayes risk only 5\% lower than the minimum Bayes risk when $\gamma = 0$. Similarly, using a power transformation offers substantial benefits in {\sc Simulation I} (41\% reduction in Bayes risk compared to the best when $\beta = 1$) and less dramatic but still notable benefits in {\sc Simulation II} (9\% reduction in Bayes risk).

The optimal $\balp$-transformed cosine dissimilarity offers a substantial reduction in Bayes risk when compared to squared Euclidean distances on the untransformed counts. In {\sc Simulation I}, the optimal $\balp^*$ resulted in a 99\% reduction in Bayes risk. In {\sc Simulation II}, the reduction was 89\%. 

Together, the simulation results under the two optimality criteria suggest that substantial gains can be achieved by choosing  $0.3 <\beta<0.5$ and  $0.4 < \gamma <0.5$, approximately.  Although they do not agree precisely on the optimal choice of $(\beta, \gamma)$, the simulations are unanimous in their exclusion of Euclidean distance and the untransformed cosine dissimilarity (i.e., $(\beta,\gamma) = (1, 0)$). They also highlight the advantage of adding  fractional pseudo counts when using the Hellinger distance: $\gamma$  should be chosen to be positive when $\beta = 1/2$. Based on these results we compare $(\beta,\gamma) = (0.3, 0.5)$ and $(0.5, 0.5)$ in the data analyses in Section~\ref{sec:data}. 


\section{Application to AIA data}
\label{sec:data}

We apply our method to the set of six SDO/AIA solar images depicted in Figure~\ref{fig:data}. These images were collected on 26 February 2015 at 20:57 UT using the AIA filters centered on 94\AA, 131\AA, 171\AA, 193\AA, 211\AA, and 335\AA.  The filters are designed to capture the most prominent features in the coronal spectrum when observing plasma in the temperature range of $10^6$--$10^7$ Kelvin.  Particular ions in plasmas of this temperature produce significant electromagnetic emission in narrow wavelength ranges. The filters capture such spectral features, in particular those resulting from Fe\,XVIII, Fe\,VIII/Fe\,XXI, Fe\,IX, Fe\,XII/Fe\,XXIV, Fe\,XIV, and Fe\,XVI, respectively.

The images in Figure~\ref{fig:data} feature a prominent coronal hole (CH) near the southern pole of the Sun. A CH is an area of mostly open magnetic field lines where the solar wind originates. Generally speaking, these regions are thought to have less variation in thermal properties than, for example, the active regions around Sun spots. This CH has many bright points within it that appear to be due to low-lying closed loops, surrounded by low-brightness, uniform regions. It is bordered by areas of significant activity, characterized by prominent loop structures. The limb of the Sun is also visible in this field, bordering the CH at the bottom of the image. While the CH appears to extend beyond the visual horizon of the limb, loops from the CH border areas are visible sticking past the limb. 

\begin{figure}[p]
	\centering
	\includegraphics[height=6.8in]{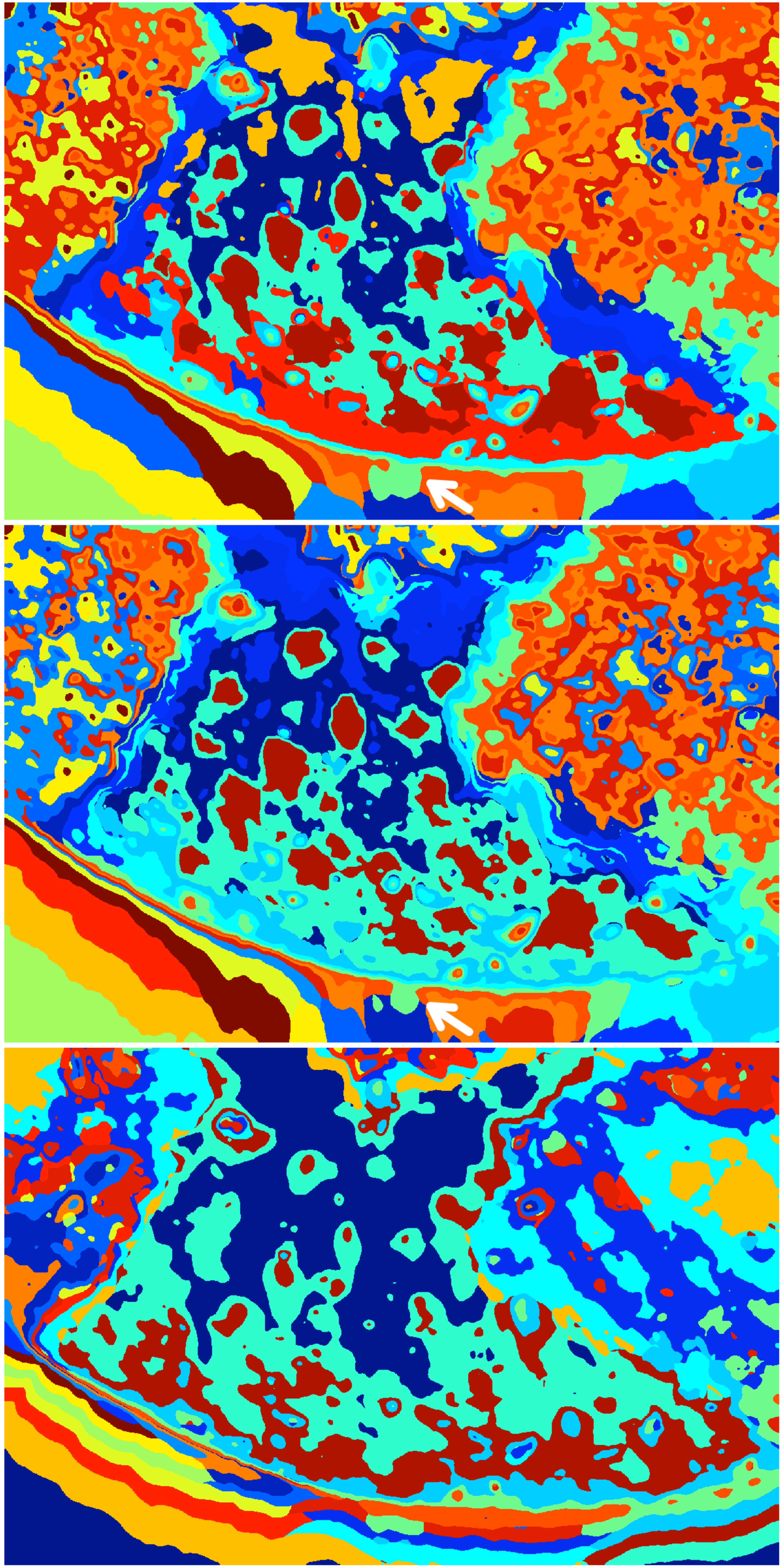}
	\caption{Segmentation results for the \edit{$1026 \times 1536$ pixel} cutout of the images in Figure~\ref{fig:data} containing the coronal hole. Segmentations were based on cosine dissimilarities with $(\beta, \gamma) = (0.3, 0.5)$ (top); cosine dissimilarities with $(\beta, \gamma) = (0.5, 0.5)$ (middle); and squared Euclidean distances between untransformed observations (bottom). (That is, the segmentation in the bottom panel is based on the off-the-shelf $k$-means algorithm.) In all cases the number of segments was set to twenty and colors are used to indicate segment labels. While there is substantial agreement between the two segmentations based on cosine dissimilarities, they both differ substantially from the $k$-means segmentation. The arrows in the top and middle panels indicate the bulb-shaped region discussed in Section~\ref{sec:secondary}.
}
	\label{fig:CH}
\end{figure}

Figure~\ref{fig:CH} displays the results of segmenting a cutout of the images in Figure~\ref{fig:data} containing the CH into twenty segments. The segmentations in the top and middle panels are based on the cosine dissimilarities in (\ref{eq:cos-transformed}) with  $(\beta,\gamma) = (0.3, 0.5)$ and $(0.5, 0.5)$, respectively. That in the bottom panel was based on squared Euclidean distances on the untransformed observations, i.e., the standard $k$-means algorithm directly applied to the observations. Prior to segmentation, we mildly spatially smoothed each image with a Gaussian filter with standard deviation $5$. \edit{This standard deviation corresponds to the approximate size of the structures of interest on the Sun, such as loops, and suppresses smaller features.} There are several evident differences between the features revealed by the cosine- and Euclidean-based segmentations. For example, in the off-limb corona at the bottom of the images, the cosine-based segmentations reveal funnel-shaped structures, while the Euclidean-based segmentation separates this region into horizontal striations. The horizontal striations are clearly attributable to simple intensity variations. Moreover, in the Euclidean-based segmentation, there is an abrupt border between the CH and the active regions surrounding it, while in the cosine-based segmentations, there appears to be a more gradual transition between the CH and its surroundings. 
We interpret these differences as an indication that potentially interesting thermal features can be disguised by off-the-shelf image segmentation techniques, in this case $k$-means.

\begin{figure}[t]
	\centering
	\includegraphics[width=\textwidth]{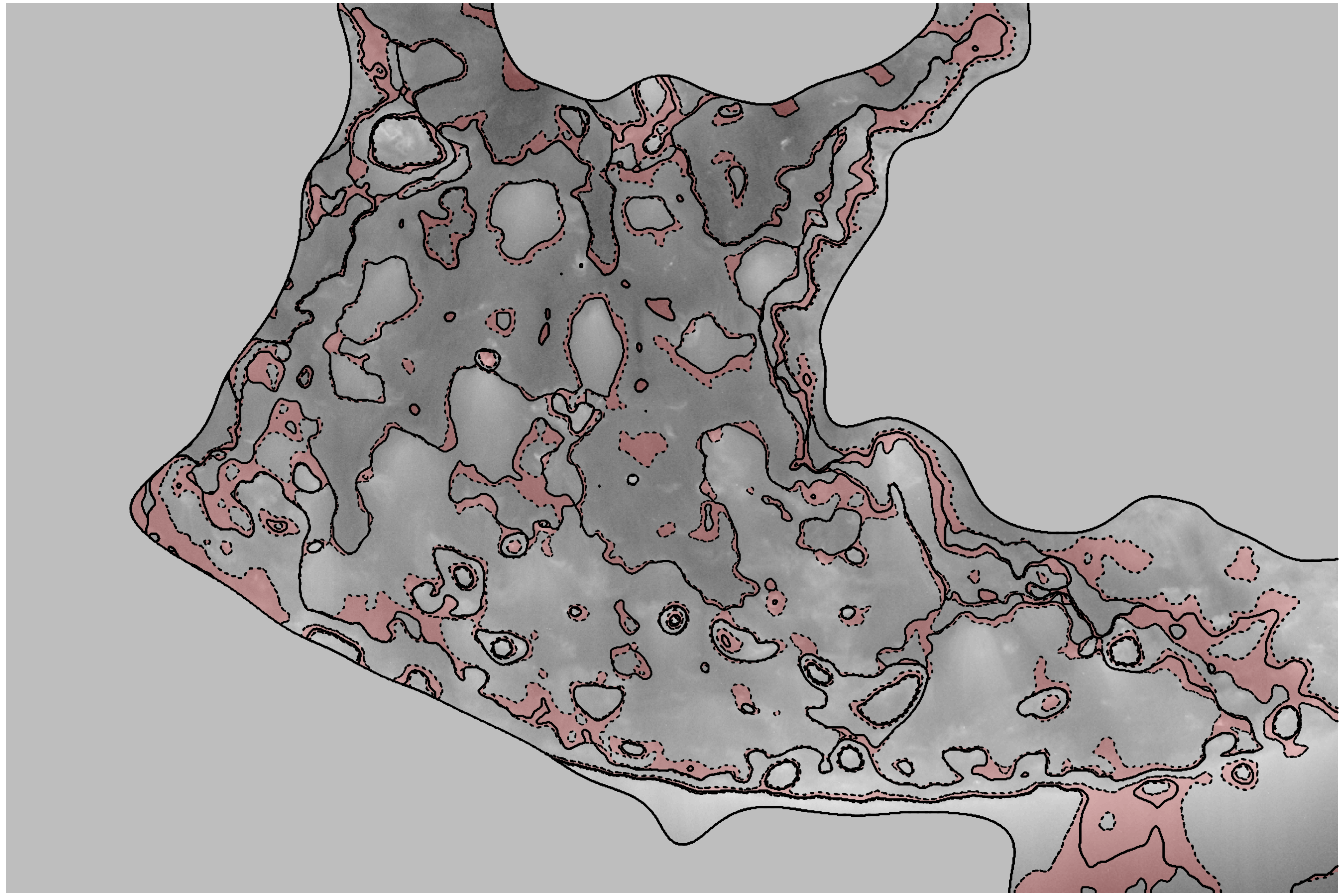}
	\caption{A comparison of the cosine-dissimilarity segmentation results  for the coronal hole region of the images in Figure~\ref{fig:data} and discussed in Section~\ref{sec:data}. Again we compare $\beta = 0.3$ (solid lines) and $\beta=0.5$ (dotted lines). Differences between the two five-cluster segmentations are highlighted in pink. (These differences were found by first relabeling the image segments to find the greatest agreement between the two segmentations, and then highlighting the pixels on which the resulting segment labels disagree.) The gray scale values in the unmasked region are determined by the intensity in the 171\AA~passband.}
	\label{fig:CH-closeup}
\end{figure}

Figure~\ref{fig:CH-closeup} provides a detailed comparison of the two segmentations based on cosine dissimilarities. In order to focus on structure within the CH and eliminate structure outside it, we created an approximate mask for the CH,\footnote{We created the mask by first smoothing the images with a 
Gaussian filter with standard deviation $30$, and then segmenting the images into two segments, using the cosine dissimilarities (\ref{eq:cos-transformed}) with $(\beta, \gamma) = (0.5, 0.5)$. We applied the mask to images smoothed with a Gaussian filter with standard deviation $5$---i.e., less smoothing than used to create the mask.}  masked the region surrounding the CH, and segmented the resulting image into five segments using the same dissimilarities as in the top and middle panels of Figure~\ref{fig:CH}. The two choices of $\beta$ lead to similar segmentations that disagree primarily on detailed features at the edges of segments. We can quantify the agreement between two segmentations by applying the adjusted Rand index to the partitions corresponding to each segmentation; there is no need for one of the partitions to be ground truth. The segments based on the two values of $\beta$ are in strong agreement, as evidence by an adjusted Rand index of $0.91$. This supports the conclusion that within the recommended range of $\beta$, the segmentations are not extremely sensitive to the particular choice of $\beta$.

From a practical point of view, Figure~\ref{fig:CH-closeup} illustrates a fast method for enhancing the contrast of solar images by obtaining segmentations that highlight spectral/thermal information rather than  intensity as do the standard brightness maps 
in Figure~\ref{fig:data}. 
We emphasize that spatial cohesion of the segments 
in Figure~\ref{fig:CH-closeup} is not imposed by our method (except insofar as we spatially smooth the observed images prior to segmentation)---we do not account for pixel locations. Rather, spatial cohesion stems from the spectral (and hence thermal) similarity of nearby pixels.

\section{Discussion}
\label{sec:disc}

\subsection{Secondary science-based statistical analyses}
\label{sec:secondary}


Our ultimate goal is to feed thermally-segmented solar images into a secondary statistical analysis. While a full exploration of such secondary analyses is beyond the scope of this paper, there are hints as to the power of this approach. The arrows in the first two panels of Figure~\ref{fig:CH}, for example, identify a bulb-shaped region of seemingly high-temperature plasma that is on first inspection predictive of a reconnection event. These occur when magnetic fields under stress reconnect and release energy in the form of flares, coronal mass ejections, or plasma jets. Such events have the potential to adversely affect the near-Earth space-based infrastructure. Thus, their prediction is a high scientific priority. Inspection of the CH over time reveals that the feature highlighted in Figure~\ref{fig:CH} was indeed the precursor of a plasma jet. 
An eruption becomes visible about one half hour after the data analyzed in Figure~\ref{fig:CH} were recorded, at $\sim$21:20 UT on 26 February 2015. 
Clearly this single event does not establish the predictive power of our segmentation, but it does highlight its potential and motivate further investigation. 

The most common current approach is to estimate the DEM for each pixel in the the image using one of the methods described in Section~\ref{sec:solar}, and, from these estimates to {\it visually} identify regions with similar thermal profiles.  This pixel-by-pixel approach is computationally costly even for the faster methods, both in time and in disk space, taking about 3--5 hours on a 2.5GHz machine to estimate each pixel-specific DEM in a $1024\times1024$  \edit{cutout of the full} image. In practice, standard errors are also needed to compare and ultimately cluster the fits. Errors are computed via parametric bootstrap with many tens of bootstrap replicates needed for each pixel so that the computational expense quickly multiplies.  
 
Our approach reverses these steps; we first cluster pixels using their observed passbands and a dissimilarity measure tuned to preserve latent thermal structure. This computation is fast, taking only about two minutes\footnote{This computation time can be reduced. For example, using the mini-batch $k$-means algorithm of \citet{scul:10}, we obtained qualitatively similar segmentations to those in Figure~\ref{fig:CH} in under 5 seconds each.} on a 3.7GHz machine to produce each image in Figure~\ref{fig:CH}, including running the $k$-means algorithm five separate times with random initial cluster centroids and choosing the best of the five resulting clusterings. More importantly, we need only estimate the DEM for each of $\sim$$10$ image segments (or as many as we can afford computationally) rather than for each \edit{of as many as $\sim$$10^7$ pixels (in a full $4096\times4096$ image)}. Even if we use the most principled available Bayesian method with MCMC fitting \citep[i.e.,][]{kash:drak:98} the computational gain will be substantial. 
  
 \edit{Our approach does not use  spatial information, except indirectly by smoothing the images prior to segmentation. Incorporating spatial structure in a model-based approach would be an interesting, though potentially computationally costly, direction for future work.}


Alternative approaches are being explored by other teams. For example, the SDO/AIA team is developing a method that uses approximate, but fast, pixel-wise thermal representations by using a pre-computed dictionary of (three temperature band) DEMs associated with statistically distinct clusters of passband intensity values (M. Weber, private communication).  This amounts to binning each of the passband intensities and associating a fitted DEM with each six-way combination of  filter-intensity bins. New DEM inversion solutions only need to be computed when previously unobserved bin combinations are encountered.  In principle, this strategy has the potential to keep up with the cadence of AIA images by generating a 4-megapixel 3-temperature-band DEM map every ten seconds.  While this allows for visual identification and inspection of patterns in thermal behavior, some form of analysis such as clustering must still be carried out on the DEM map to objectively identify regions that are thermally similar. It is also possible that by binning the DEMs into only three temperature bands, some thermal features obtainable from the six-band filter intensities could be lost. Because this dictionary-based method is still in development, we must leave such questions to future investigations.


\subsection{Preprocessing data for science-driven analyses}
Today's state-of-the-art astronomical data are of exceptional quality, composed of diverse and sometimes massive data streams, and are often tailored to specific scientific goals.  This ``big data'' is not just ``big,'' however; it is rich, deep, and intricate, encompassing, for example, high resolution spectrography and imaging across the electromagnetic spectrum; the intricate measurement of stellar wobble used to identify exoplanets and estimate their masses, distances and orbital shapes; and incredibly detailed movies of the dynamic and explosive processes in the solar atmosphere.  These studies aim to improve our  understanding of the evolution of the Universe and of our own origins. Such ambitious goals require descriptive science-driven statistical models and methods that relate our best understanding of underlying physical processes to observables. In the long run, this will improve understanding of the underlying processes, inform future data collection to maximize their information content, and enable further advances. Methods that simply identify patterns in data are not well-suited to such problems. Instead, we require methods that incorporate understanding of the underlying physical processes. We refer  to general-purpose methods that aim to identify patterns and clusters in data and that are generally used for prediction as  {\it data-driven} methods.  {\it Science-driven} methods, on the other hand, are typically designed for a specific inference problem and, in astronomy,  incorporate physics-based models that enable scientifically meaningful statistical inference.

While there is no general formula for the implementation of science-based methods in a big-data environment, some patterns are emerging.   In astrophysics and solar physics, a focus on the underlying physics must always be the guiding principal for methodological development. 
To be scalable, however,  methods require some sort of data filtering---a reduction in data volume that insofar as possible maintains information content. In practice, this often takes the form of preprocessing the data. In this article, we segment solar images in anticipation of secondary analyses that investigate thermal properties of the resulting segments and/or model their evolution. Similar examples abound.  The Large Hadron Collider at CERN, for instance, produces millions of proton collisions per second \citep[e.g.,][]{vand:14}. The experiments involved in the discovery of the Higgs Boson have fast triggers that make decisions about which events are worth saving, secondary analyses aim to further reduce background, and finally a science-driven analysis is employed to identify excess particles at the fitted Higgs mass.  {Data-driven} and {science-driven} methods are combined by implementing a sequence of discrete analyses: the first analyses filter the data, and latter analyses aim to answer specific scientific questions.  In this paper, we are able to embed science-driven methods in the initial data reduction phase. Others have employed similar strategies. \citet{sten:etal:13}, for example, used mathematical morphology \citep[e.g.,][]{Soille03} to efficiently identify and summarize scientifically meaningful features in solar images of  active regions.  The result is a concise numerical summary of the complexity of a magnetic flux distribution that (i) is far easier to work with than the source images, (ii) efficiently encapsulates scientifically relevant information, and (iii) is amenable to sophisticated follow-up statistical analyses.

Generally speaking there are significant challenges in implementing statistically coherent multiphase analyses. Building on the framework of Multiple Imputation, \citet{bloc:meng:13} develop a theoretical description of a class of multiphase scenarios. They illustrate a number of pitfalls that may arise, especially when the model used for preprocessing the data is unknown to the downstream analyst or is uncongenial with the downstream model; see also  \citet{meng:94}. 
 In principle, this is less of a problem in the current setting, at least if the several discrete phases of the analyses are conducted by the same researcher or the same team. 
This being said, the methods employed for data reduction (e.g., $k$-means for image segmentation) are not typically likelihood based and thus may not be easily integrated into an overarching probabilistically principled framework. Bridges can sometimes be built between data-driven methods and likelihood-based science-driven methods, however, in an effort to put the overall analysis on a firmer theoretical footing. \citet{lee:etal:11} and \citet{xu:etal:14}, for example, show how principal component analysis can be used to markedly reduce the dimension of replicates used to describe the uncertainty in the instrumental operating characteristic of X-ray telescopes. They then quantify the results of this data-driven analysis as a prior distribution for the unknown operating characteristics and  use it in a secondary science-driven likelihood-based analysis. 

In this article we develop another example of a multi-phase analysis that aims to implement science-driven methods in a big-data context. In particular, we develop a suite of image segmentation methods for massive streams of high-resolution solar images to map regions in the solar corona with similar thermal properties. Although our image segmentation methods are largely data-driven, they are designed with the ultimate scientific goal in mind and specifically aim to improve the efficiency of the follow-up analyses by producing summaries of the raw data that maintain relevant information and are amenable to science-based model fitting. 

\subsubsection*{Acknowledgments}

This project was conducted under the auspices of the CHASC International Astrostatistics Center. CHASC is supported by NSF grants 
DMS 1208791, DMS 1209232, DMS 1513492, DMS 1513484, and DMS 1513546. David van Dyk also acknowledges support from a Wolfson Research Merit Award provided by the British Royal Society and from a Marie-Curie Career Integration Grant provided by the European Commission and Vinay Kashyap from a NASA contract to the {\it Chandra} X-Ray Center NAS8-03060 as well as travel support for collaborative visits as part of the Indo-US Center for Astronomical Object and Feature Characterization and Classification, sponsored by the Indo-US Science and Technology Forum (IUSSTF).
In addition, we thank CHASC members for many helpful discussions.

\appendix

\section{Appendix}\label{sec:proof}

In this section, we prove (\ref{eq:cos-hell-general}). To simplify notation, let $\bz_i = (z_{i1}, \ldots, z_{ib}) = (y_{i1}+\gamma, \ldots, y_{ib}+\gamma)$, and let $\bz_i^{\beta} = (z_{i1}^{\beta}, \ldots, z_{ib}^{\beta}) = \bT(\by_i; (\beta, \gamma))$. Then
\begin{align*}
	\deuc^2\lt(\frac{\bz_i^{\beta}}{||\bz_i^{\beta}||_p}, \frac{\bz_j^{\beta}}{||\bz_j^{\beta}||_p}\rt) &=
	\sum_{l=1}^b \lt\{ \frac{z_{il}^{\beta}}{\lt(\sum_{m=1}^b z_{im}^{p \beta}\rt)^{1/p}} - \frac{z_{jl}^{\beta}}{\lt(\sum_{m=1}^b z_{jm}^{p \beta}\rt)^{1/p}} \rt\}^2 \\
	&= 	\sum_{l=1}^b \lt\{ \lt(\frac{z_{il}^{p \beta}}{\sum_{m=1}^b z_{im}^{p \beta}}\rt)^{1/p} - \lt(\frac{z_{jl}^{p \beta}}{\sum_{m=1}^b z_{jm}^{p \beta}}\rt)^{1/p} \rt\}^2 \\
	&= \deuc^2\lt\{\lt(\frac{\bz_i^{p \beta}}{||\bz_i^{p \beta}||_1}\rt)^{1/p},
	\lt(\frac{\bz_j^{p \beta}}{||\bz_j^{p \beta}||_1}\rt)^{1/p}\rt\}.
\end{align*}
Setting $p=2$ and dividing by two, we obtain (\ref{eq:cos-hell-general}). 

\bibliographystyle{natbib}
\bibliography{DEM}

\end{document}